\documentclass[preprint,11pt]{article}

\usepackage{arxiv}

\usepackage{amssymb}
\usepackage{float}
\usepackage{geometry}
\usepackage[caption = false]{subfig}
\usepackage{graphicx}
\usepackage{amsmath}
\usepackage{xcolor}
\usepackage{lineno}
\usepackage{multirow}
\usepackage{amsthm}
\usepackage{amsmath}
\usepackage{esvect}
\usepackage{hyperref}

\usepackage{algpseudocode}
\usepackage{algorithm}

\title{Automatic denoising and differentiation based on Savitzky-Golay filtering and Homogeneous Differentiators for attractor reconstruction via differential embedding}

\author{Uros Sutulovic \\
	Department of Industrial Engineering\\
	University of Trento\\
	Trento, Italy \\
	\texttt{uros.sutulovic@unitn.it} \\
\And
	Daniele Proverbio \\
	Department of Industrial Engineering\\
	University of Trento\\
	Trento, Italy \\
	\texttt{daniele.proverbio@unitn.it} \\
\And
	Rami Katz \\
	School of Electrical and Computer  Engineering\\
	Tel Aviv University\\
	Tel Aviv-Yaffo, Israel \\
	\texttt{ramkatsee@tauex.tau.ac.il} \\
\And
	Giulia Giordano \\
	Department of Industrial Engineering\\
	University of Trento\\
	Trento, Italy \\
	\texttt{giulia.giordano@unitn.it} \\
}

\renewcommand{\headeright}{}
\renewcommand{\undertitle}{}

\begin{document}

\maketitle

\begin{abstract}
Differential embedding methods aim to reconstruct attractors of dynamical systems from noisy measured time series, but require accurate estimates of signal derivatives, which are difficult to obtain in the presence of noise. We introduce SHADED (\textbf{S}avitzky-Golay and \textbf{H}omogeneous-differentiator based \textbf{A}utomatic \textbf{DE}noising and \textbf{D}ifferentiation), a novel methodology for effective denoising and estimation of derivatives up to an arbitrary order, which enables attractor reconstruction via differential embedding from noisy time series data.
Homogeneous sliding-mode Differentiators (HD) guarantee finite-time derivative estimates in the presence of noise, while subsequent Savitzky-Golay (SG) filtering attenuates chattering.
Crucially, SHADED extracts all parameters required for application of both HD and SG automatically from the data,
without requiring time-consuming manual tuning that may lead to inaccurate reconstruction,
and can also incorporate prior knowledge, if available, thereby yielding a flexible and effective tool for data-driven numerical differentiation of noisy signals; the obtained derivative estimates are valuable \textit{per se} and instrumental for attractor reconstruction. 
The obtained differential embeddings can reveal features of the underlying dynamics that are useful, \textit{e.g.}, for system identification, pattern recognition and discrimination between dynamic regimes; the latter application is particularly important in biomedical settings, to help distinguish between different physiological and pathological states of a patient.
We demonstrate the efficacy of SHADED by testing it on computational neuroscience models, LTspice-simulated chaotic electronic circuits, and photoplethysmography and arterial blood pressure experimental recordings (where the base signal and noise are unknown):
across all these case studies, under varying signal-to-noise ratios and different noise types, SHADED produces accurate derivative estimates and accurate attractor reconstructions via differential embedding (whenever a ground truth is available) or geometrically coherent and reproducible reconstructions consistent with the expected dynamics (in the absence of a ground truth), with short computation times, and \emph{without the need for manual parameter tuning}.
\end{abstract}


\section{Introduction}
\label{sec:introduction}

\subsection{Attractor reconstruction from noisy time series}
Many problems in nonlinear science involve the estimation of dynamical attractors from 
noisy time-series data generated by an unknown dynamical system. Usually, only a small subset of the system states are measured, and such measurements are affected by noise \cite{bradley2015nonlinear}, thereby providing only partial and indirect information on the underlying dynamics. In such cases, the desired properties of the system must be inferred from the time series via data-driven signal processing techniques. In this work, we consider noisy time-series data generated by (unknown) finite-dimensional systems modelled by nonlinear ordinary differential equations of the form
\begin{equation}
  \dot{\boldsymbol{x}}(t)=f(\boldsymbol{x}(t)), \quad \bar{y}(t)=h(\boldsymbol{x}(t)), \quad \boldsymbol{x}(t)\in \mathcal{X}\subseteq \mathbb{R}^n,
  \label{eq:generic_ode}
\end{equation}
where $\boldsymbol{x}(t)$ is the state vector, $\mathcal{X}$ is the state space, $f \colon \mathbb{R}^n\rightarrow\mathbb{R}^n$ is a smooth vector field,
and $h \colon \mathbb{R}^n\rightarrow\mathbb{R}$ is a smooth measurement function producing the scalar observable $\bar{y}(t)$. 
In applications, one does not observe $\bar{y}(t)$ directly, but only noisy discrete-time samples of it. 
Our focus is on recovering, from such noisy scalar observations of \eqref{eq:generic_ode}, the attractors associated with the system.  
An attractor is an invariant set in $\mathcal{X}$ that is asymptotically approached by some or all trajectories after transient behaviour has decayed \cite{milnor1985concept}. 
Attractors capture the long-term behaviour of the dynamical system, including convergence to a fixed point, limit-cycle oscillations, or chaotic motion \cite{ott1993chaos}, and are used to geometrically visualise the system regimes. Different dynamical regimes are associated either with  different attractors or with different values of numerical quantities such as fractal dimensions, Lyapunov exponents, entropy-based indices, or recurrence statistics \cite{bradley2015nonlinear,grassberger1983characterization}. 
Reconstructed attractors have been used in cardiovascular and electroencephalogram (EEG) applications to quantify signal morphology and to distinguish between dynamical regimes associated with physiological and pathological states \cite{nandi2018novel,zhang2024periodic,pourdavood2024eeg}. 
Accurate reconstruction of a desired attractor, up to a choice of coordinates on $\mathcal{X}$, is therefore highly desirable for system identification and classification tasks \cite{sauer1991embedology,lee2013smooth}. Reconstructed attractors are also fundamental for subsequent analyses: persistent homology and related tools from topological data analysis can be applied directly to reconstructed attractors to obtain robust topological invariants that are useful for comparing dynamical regimes and detecting qualitative changes in system behaviour \cite{venkataraman2016persistent,lucas2025topological}. 
Similarly, reconstructed phase-space representations have been used to derive topological early-warning measures for critical transitions and bifurcations \cite{syed2021using}. 
In all such cases, the resulting quantities are meaningful only if the underlying attractor reconstruction is sufficiently accurate to contain information on the original dynamics: trajectory distortions introduced by finite sampling, measurement noise, or poorly chosen reconstruction parameters can propagate into the computed quantities and lead to misleading conclusions about the system and its regime \cite{kennel1992determining}. 

\subsection{Existing work}

Existing literature on attractor reconstruction can be broadly grouped into three categories: \textbf{(i)} methodological work devising attractor reconstruction techniques; \textbf{(ii)} applications that use existing reconstruction techniques on specific datasets; and \textbf{(iii)} works that both introduce methodological innovations and demonstrate them on application data.

\textbf{(i)} Methodological work on attractor reconstruction is mostly centered on time-delay embedding, which maps a scalar time series to a collection of points obtained from multiple shifted (delayed) copies of the data. 
The theoretical basis for these methods was established by Takens \cite{takens2006detecting} and expanded by Sauer, Yorke, and Casdagli \cite{sauer1991embedology}, who showed that, under suitable conditions, a time series can be used to reconstruct the attractor up to a smooth change of coordinates (continuously differentiable map with a continuously differentiable inverse), provided that the embedding dimension, i.e. the number of coordinates used to represent the attractor, is sufficiently large. 
The resulting reconstruction technique must be finely tuned to guarantee correct attractor reconstruction: the time series should be sampled so as to ensure signal variation that contains sufficient information about the attractor, and the embedding dimension must be chosen carefully. 
If the dimension is too small, distinct regions of the attractor are folded together, and, if it is too large, the reconstruction becomes more sensitive to noise in the data, which can obscure the underlying dynamics \cite{bradley2015nonlinear,kennel1992determining}. 
Thus, time-delay embeddings inherently require judicious parameter selection for the time delay and the embedding dimension.
Common heuristics use the first minimum of the average mutual information to select a delay at which successive coordinates are as independent as possible \cite{fraser1986independent}, and false-nearest-neighbour analysis to select an embedding dimension at which projection-induced overlaps become negligible \cite{kennel1992determining}. 

In real applications, however, the ideal assumptions underlying the embedding theorems (noise-free signals, arbitrarily long time series, and the availability of informative measurement functions) are rarely satisfied \cite{bradley2015nonlinear}. 
Recordings are usually finite and quantised, and they are corrupted by noise whose distribution and temporal structure are unknown. 
When the ideal assumptions do not hold, the theoretical guarantees of the embedding theorems are lost and the reconstruction bottleneck becomes selecting parameters that yield, from the available data, an attractor representation judged sufficiently informative and robust for the intended analysis.
This has motivated extensive methodological work on parameter selection within the delay-embedding framework \cite{kennel1992determining,fraser1986independent,tan2023selecting,huang2024detecting}, which can substantially improve reconstruction quality compared with naive parameter choices.
Because time-delay embeddings are implemented directly on noisy data and reconstruction quality is highly sensitive to parameter choices, denoising has also been incorporated into reconstruction procedures. 
The Schreiber-Grassberger method \cite{grassberger1993noise}, widely used in the literature, denoises a delay-coordinate reconstruction by iteratively replacing each point with a local average of its nearest neighbours. 
This procedure reduces the effects of measurement noise and numerical artifacts in the reconstructed trajectory, while preserving the large-scale geometry of the attractor.
In practice, this method also depends on parameters such as the neighbourhood size and the number of iterations, whose tuning is nontrivial in the absence of a ground truth. 
Ultimately, in realistic scenarios, both denoising and parameter selection become critical for obtaining a reliable reconstruction. 
As a result, automatic and data-driven parameter selection strategies that can operate well in the presence of noisy measurements are extremely sought after in applications.

Differential embeddings provide an alternative reconstruction strategy. 
Under appropriate assumptions, Takens showed that one can reconstruct the attractor by using successive time derivatives of a time series instead of its delayed copies \cite{takens2006detecting}. Such a reconstruction is more informative than time-delay embeddings, since derivative-based coordinates can be more directly tied to the local evolution of the system, via the notion of observability \cite{letellier2005relation}, and are therefore more useful for regime discrimination and model identification. 
However, although differential embeddings avoid some of the limitations associated with time-delay embeddings, they introduce a distinct challenge: greater sensitivity to measurement noise.
In fact, numerical differentiation amplifies noise, whence accurate derivative estimation is a particularly central challenge \cite{letellier2005relation}. 
\emph{Accurate numerical differentiation in the presence of noise} is thus a key component of any differential embedding approach.

Previous work has investigated numerical differentiation methods for differential embeddings. 
Finite-difference schemes are simple and computationally inexpensive, but they are strongly noise-amplifying, especially for higher-order derivatives \cite{chartrand2011numerical,knowles2014methods}. 
Regularisation-based methods can reduce this sensitivity by enforcing smoothness or sparsity, but they require careful choice of penalty terms and regularisation parameters \cite{chartrand2011numerical,knowles2014methods}, further exacerbating the parameter selection problem. 
Algebraic differentiators offer another compromise by using finite-window polynomial approximations or related integral formulas, although their performance depends strongly on the window length, polynomial order, and smoothness assumptions on the signal \cite{li2020exploring,komarov2025taxonomy}. 
Homogeneous Differentiators (HD) approximate derivatives through nonlinear sliding-mode systems and can achieve finite-time convergence under appropriate assumptions on signal smoothness and noise properties \cite{levant2003higher,levant2020robust}. 
Their main strengths are the robust estimation of derivatives of arbitrary order in finite time (which circumvents the need for extremely long time series) and the existence of rigorous estimation guarantees for certain classes of noise. However, their practical performance still depends on parameter (gain) selection and discretisation \cite{hanan2021low}. 
In particular, inappropriate gain choices can lead to chattering, \textit{i.e.}, rapid oscillations in the output, whose amplitude increases with the gain.
Nevertheless, for differential embeddings, HD-based techniques possess a promising combination of accuracy and noise robustness \cite{sutulovic2025efficient}, provided one can resolve the practical challenges in implementation, which are related to the problem of parameter selection.
Ideally, parameters need to be selected based on the available noisy time series alone, as both the underlying dynamic model of the studied phenomenon and noise-free measurements are unavailable: this motivates the present work.

It is worth stressing that reliable derivative estimates are valuable \textit{per se}, beyond attractor reconstruction. For instance, in model-discovery and physics-informed learning frameworks, derivatives are used to build candidate libraries of dynamical terms and enforce consistency with governing equations \cite{brunton2016discovering,champion2020unified,wang2016data}. 
These applications lie outside the scope of this paper, but reinforce the importance of robust derivative estimation, by showing how it enables system identification, attractor reconstruction and downstream dynamical inference when derivatives are estimated reliably.

\textbf{(ii)} Application-driven studies form a second broad category of existing works, in which reconstructed attractors and their geometric properties are used to characterise waveform shape and variability and to distinguish dynamical regimes in specific applications. 
In cardiovascular physiology, attractor reconstructions of arterial pulse waveforms have been used to quantify waveform morphology through attractor features such as attractor size, rotation angle, shape, and trajectory density \cite{nandi2018novel,aston2018beyond,horandtner2022attractor}. 
These quantities have been applied to haemodynamic monitoring \cite{nandi2018novel,aston2018beyond}, assessment of slow-breathing interventions \cite{horandtner2022attractor}, photoplethysmography (PPG) signal-quality evaluation \cite{schmith2023photoplethysmography}, age-related analysis of fingertip PPG dynamics \cite{sun2026analysis}, and beat detection in PPG \cite{pettit2024photoplethysmogram}. 
In these studies, reconstruction is typically performed with time-delay embeddings and the main emphasis is on how to extract 
quantitative features and the recurrence structure of the reconstructed attractor.
For electrocardiogram (ECG) analysis, the authors in \cite{noponen2009invariant} proposed invariant trajectory-classification methods employing reconstructed trajectories to distinguish cardiac conditions while remaining robust to changes in waveform morphology caused by posture or respiration.
In EEG analysis, \cite{pourdavood2024eeg} introduced spectral attractors by combining band-pass filtering, principal component analysis, and time-delay embedding. 
Their method builds a low-dimensional representation of the EEG signal in which recurrent patterns of activity become visible, making it possible to track how dynamics evolve across standard EEG bands, and to reveal age-related changes not visible in conventional spectral measures. 
Attractor-based measures have also been used in epilepsy research, where changes in reconstructed EEG attractors during seizure onset provide a dynamical signature of ictogenesis \cite{quyen2003toward}. In these works, the reconstruction step is aimed at obtaining a consistent and informative geometric representation of the dynamics, from which various secondary indicators and task-specific features can be derived. However, most of these works rely on existing reconstruction techniques with parameters chosen manually or through heuristics tailored to the problem, and they seldom analyse how noise and parameter variability affect the extracted features and how the presence of noise and variability in the data should inform
the choice of parameters
for the employed algorithms.

\textbf{(iii)} In a third category of contributions, specialised reconstruction-related techniques are developed or extended and then validated on data, often as part of a broader dynamical analysis. 
A recent example is \cite{kramer2021unified}, which generalises time-delay embeddings by combining non-uniform delays with a cost function based on continuity (nearby states should map to nearby images) and prediction-error statistics (how well short-term forecasts from the reconstructed space match the observed data), and provides a fully automatic procedure for selecting delays and embedding dimensions from univariate or multivariate time series. 
The method is evaluated on chaotic systems and experimental data from chaotic chemical oscillators, illustrating how an automatic delay-based reconstruction can improve downstream analyses. 
However, it is confined within the delay-embedding paradigm, therefore still requiring the selection of informative delays and embedding dimensions from the observed data, and it does not include an explicit denoising stage for the raw time series; noise is only handled implicitly through the continuity and prediction-error criteria, which may result in meaningless reconstructed attractors if the noise is sufficiently strong (see Supplementary Fig. S17).
Inherently different approaches aim at reconstructing attractors from noisy observations while relying on derivative information extracted from the data.
In particular, our related work \cite{sutulovic2025efficient} combined, for the first time, Homogeneous Differentiators (HD) with Savitzky-Golay (SG) filtering to build \emph{differential} embeddings from noisy signals, yielding faithful attractor reconstruction; the method proposed therein was applied to simulated systems as well as to zebrafish neurological data for the study of seizure dynamics, and the reconstructed attractors were shown to preserve accuracy even under challenging additive and multiplicative noise conditions. 
However, the approach relied on manual tuning of the HD gains and SG window sizes for each dataset, thereby limiting its practicality for attractor reconstruction, because manual tuning is time consuming and accurate tuning criteria cannot be determined without knowing the true noise-free dynamics.

Here, we address the open parameter selection problem that remained unresolved in \cite{sutulovic2025efficient} and that also poses a very general challenge: in fact, the automatic selection of parameters for Savitzky-Golay filters \cite{krishnan2012selection,sadeghi2020window} and Homogeneous Differentiators \cite{wang2025practical,tonti2023optimal} is a long-standing open problem. 
We introduce SHADED, a general automatic HD-SG-based method for denoising and derivative estimation, which enables attractor reconstruction via both time-delay and differential embedding, where all required HD gains and SG window sizes are \emph{inferred directly from data}, thereby avoiding the need for manual parameter selection, while achieving high reconstruction accuracy.

\subsection{Contribution}
\label{sec:intro_contribution}

When one seeks a widely applicable methodology for numerical differentiation and attractor reconstruction, manual tuning of HD gains and SG window sizes, which are crucial parameters, is impractical.
To solve this issue, we develop SHADED (\textbf{S}avitzky-Golay and \textbf{H}omogeneous-differentiator based \textbf{A}utomatic \textbf{DE}noising and \textbf{D}ifferentiation), a fully automatic methodology for denoising and estimating successive time derivatives, which thus enables attractor reconstruction also via differential embedding. 
SHADED is built around a systematic data-driven procedure for selecting the HD gain and the SG window size.
Addressing such parameter selection problems is of interest \textit{per se} \cite{krishnan2012selection,sadeghi2020window,wang2025practical,tonti2023optimal} and is instrumental to ensure that derivatives of arbitrary order, and hence differential embeddings, can be obtained reliably from noisy data, without requiring manual intervention, also when the system dynamics, the ground truth signal and the noise properties are unknown.

Conceptually, SHADED consists of a staircase architecture composed of efficient modules, each comprised of an HD followed by an SG filter. 
The first module takes the noisy measurement as input and produces a denoised signal estimate; subsequent modules operate on the output of the previous one in the staircase to estimate higher-order derivatives, up to a prescribed order. 
At each level, both the HD gain and the SG window size are selected automatically from the data using criteria based on the residual signal, i.e. the difference between the input and output of the corresponding module; optional user tuning is possible when prior information on the signal or the noise level is available. 
Given a noisy signal, the overall procedure automatically yields denoised derivatives up to the desired order, which can then be used for attractor reconstruction via differential embedding. 
Fig.~\ref{fig:methodology_pipeline} shows why more than two parameters are required: each derivative order has its own HD gain and SG window size, since trade-offs between noise sensitivity and preservation of the low-frequency signal content change between the estimated derivatives.
SHADED determines all the required parameters (two for each derivative) in an automated way; a more in-depth description of the method 
is provided in Section~\ref{sec:methodology}.

The present work makes three essential contributions:
\begin{itemize}
\item For automatic HD gain selection, we formulate an optimisation problem that selects the gain based on the noisy data, by analysing the difference between the differentiator output and the noisy input signal. Considering synthetic examples where a noise-free reference is available, we show numerically  that the proposed gain selection procedure yields derivative estimates whose error is comparable to derivative estimates obtained via manual parameter tuning based on the noise-free signal (which is however inaccessible in practice in concrete applications).
\item For automatic SG window-size selection, we formulate an optimisation problem that balances chattering suppression against preservation of the low-frequency signal content, under a spectral-separation assumption. 
Specifically, we assume that the chattering content is concentrated at high frequencies and we seek the minimal window size that removes the frequency components associated with chattering while retaining the low-frequency content of the signal.
The candidate window obtained from this optimisation can then be used directly, or optionally adjusted by employing a frequency-domain persistence criterion applied to the difference between the corresponding SG filter input and output signals. The final SG window is shown to successfully attenuate chattering, while avoiding oversmoothing of the most informative signal components for the geometry of the reconstructed attractor.
\item We integrate both automatic procedures into SHADED, a recursive architecture for high-order derivative estimation, and demonstrate that the automatically estimated derivatives allow for differential embeddings that preserve the geometry of the underlying attractor across multiple reconstruction problems, including theoretical neuronal models, transistor-based circuits and empirical cardiovascular waveforms; see Section~\ref{sec:results} and Supplementary Section S2 for details. 
\end{itemize}

SHADED
builds upon 
the HD-SG approach first introduced in \cite{sutulovic2025efficient}
and provides 
a fully automatic, data-driven procedure for parameter selection, thereby making HD-SG-based differential embeddings a practical tool for real data applications.

On the one hand, the obtained denoised derivative estimates of the signal can be useful in their own right in numerous applications; on the other hand, the attractors reconstructed from noisy measurements by SHADED can serve as a basis for subsequent computations, such as topological data analysis, system identification, classification or pattern-recognition. 

The paper is structured as follows.

Section~\ref{sec:methodology} presents SHADED, the developed computational methodology consisting of a recursive HD-SG algorithm for higher-order derivative estimation, where HD gain and SG window size are automatically selected from the noisy measurements. 
Section~\ref{sec:results} evaluates the performance of SHADED on theoretical neuronal models, noisy transistor circuits, and empirical cardiovascular recordings, demonstrating the quality of the derivative estimates obtained through automatic tuning and illustrating how the resulting attractors, reconstructed via differential embedding, recover the geometry of the underlying dynamics. 
Practical aspects, limitations, and directions for future work are discussed in Section~\ref{sec:conclusions}, while detailed properties of HD, together with additional numerical results, are reported in the Supplementary Material and are referred to throughout the manuscript.

\subsection{Notation}
\label{sec:notation}
Given $N\in \mathbb{N}$, we denote $[N]=\left\{1,\dots,N \right\}$ and $[N]_0=\left\{0,1,\dots,N \right\}$.
Bold symbols, such as $\boldsymbol{x}$, denote vectors, whereas $\boldsymbol{x}(t) \in \mathbb{R}^n$ denotes the (inaccessible) state of a continuous-time dynamical system given by \eqref{eq:generic_ode}. 
The noise-free observation is denoted $\bar{y}(t)=h(\boldsymbol{x}(t))$, as in \eqref{eq:generic_ode}, whereas the actual 
measured signal is denoted by $y(t)$ and is assumed to be affected by either additive or multiplicative noise, i.e.  $y(t)=\bar{y}(t)+\eta(t)$ or $y(t)=(1+\eta(t))\bar{y}(t)$, respectively,
where $\eta(t)$ denotes the noise corrupting the measurements; the considered noise types are presented in Section~\ref{sec:results}.

For a continuous-time signal $g(t)$ sampled at uniform sampling times $t_k = k\Delta$, $k\in [K]_0$, with sampling period $\Delta>0$, we denote the corresponding discrete-time sequence by $\{g_k\}_{k=0}^K$, where $g_k=g(t_k)$.
Although the examples in this paper are based on uniformly sampled data for simplicity, SHADED \emph{does not} strictly require uniform sampling: the discrete-time HD formulation extends to non-uniform sampling grids, and the SG windows can be defined in terms of time neighbourhoods instead of a fixed sample count (see Section~\ref{sec:conclusions} for further details).

Throughout the paper, hats denote quantities that are selected automatically by the algorithms introduced in Section~\ref{sec:methodology}, while stars denote the corresponding reference values obtained by minimising the root-mean-square error (RMSE) with respect to the noise-free signal, when such a reference is available. 
For example, $\hat{L}$ and $\hat{w}$ ($\hat{T}$) are the automatically selected HD gain and SG filter window size (window time span), respectively, whereas $L^\star$ and $w^\star$ ($T^\star$) denote the values that minimise the RMSE relative to the noise-free signal $\bar{y}(t)$.

\section{Methodology}
\label{sec:methodology}

\begin{figure}[ht!] 
\centering\includegraphics[width=1\linewidth]{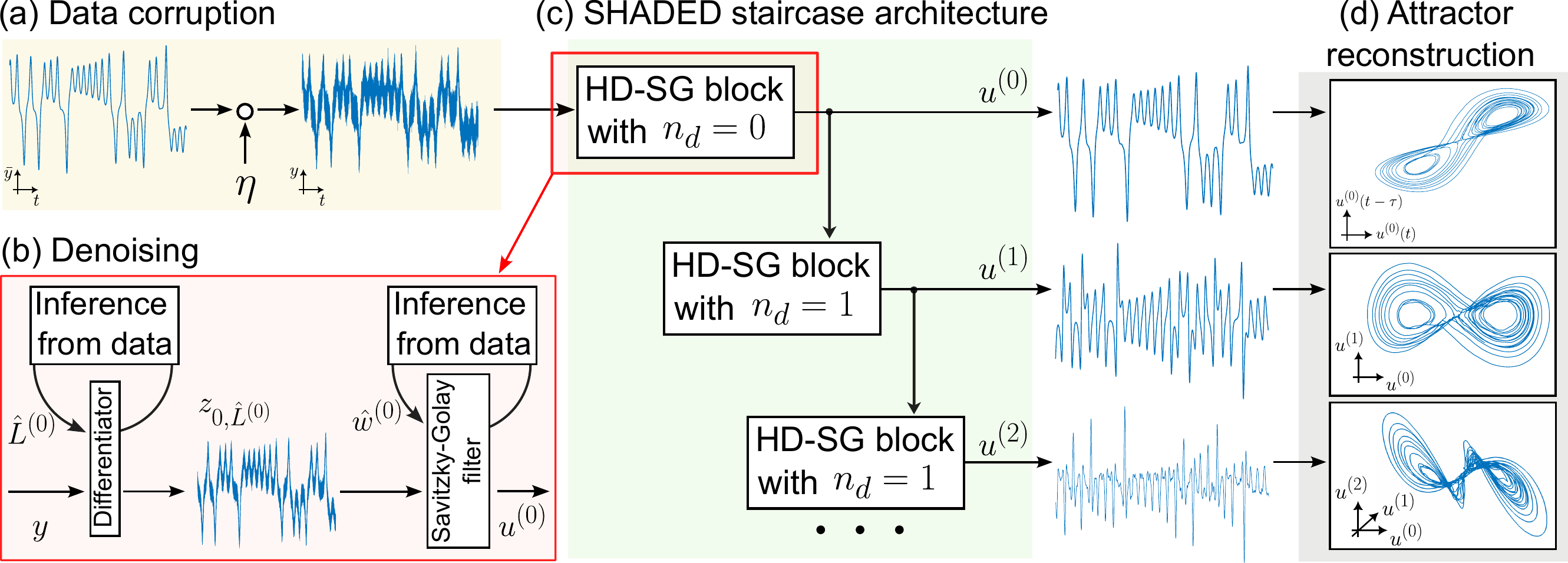}
\caption{\textbf{Given a noisy scalar measurement, SHADED automatically yields its denoised derivatives, which can be used for attractor reconstruction, with both time-delay and differential embedding.}
We demonstrate the methodology on the noisy output of a Lorenz system (see Supplementary Table S1).
(a) The noise-free signal $\bar{y}$ is corrupted by measurement noise $\eta$, yielding the noisy measurement $y$; here, $y=\bar{y}+\eta$ with $\eta \sim \mathcal{N}(0,5)$.
(b) The noisy measurement $y$ is processed by a first HD-SG block with differentiation order $n_d=0$. The HD gain $\hat{L}^{(0)}$ is automatically estimated by minimising $\mathcal{C}_{\mathrm{diff}}(L)$ in \eqref{eq:diff_cost} (see also Fig.~\ref{fig:diff_autotuning}), yielding the signal $z_{0,\hat L^{(0)}}$, which is likely affected by chattering. 
To attenuate chattering, the SG window size $\hat{w}^{(0)}$ is automatically chosen by minimising $\mathcal{C}_{\mathrm{SG}}(w)$ in \eqref{eq:sg_cost} and then possibly applying the persistence criterion described in Section~\ref{sec:freq_persistence} (see also Fig.~\ref{fig:SG_autotuning}), thus yielding the denoised signal $u^{(0)}$.
(c) Estimates of higher order derivatives $u^{(1)},\ldots,u^{(N)}$ are obtained by recursively applying HD-SG blocks to the output of the previous step: at each level, the HD with automatically tuned gain $\hat{L}^{(\ell)}$ and differentiation order $n_d=1$ estimates the next derivative, and the SG filter with tuned window $\hat{w}^{(\ell)}$ attenuates chattering while limiting suppression of the low-frequency content of the signal.
(d) The obtained time series of derivative estimates $\{u^{(\ell)}\}_{\ell=0}^{N}$ can then be used to construct either a time-delay embedding from $u^{(0)}$ (2D case shown in the first row, see also Supplementary Fig. S17) or a differential embedding using $\{u^{(\ell)}\}_{\ell=0}^{N}$ (cases $N=1$ and $N=2$ respectively shown in the second and third rows).}
\label{fig:methodology_pipeline}
\end{figure}

We can see SHADED as partitioned into constituent modules, according to the flow illustrated in Fig.~\ref{fig:methodology_pipeline}. 
Here, we describe in detail the two automatic tuning procedures that underpin the iterative HD-SG staircase, namely the residual-based selection of the HD gain and the SG window-size selection based on chattering and signal-content preservation metrics, highlighted in the red box in Fig.~\ref{fig:methodology_pipeline}(b).  Fig.~\ref{fig:methodology_pipeline}(c) shows how these blocks are recursively assembled into the full staircase architecture. 

\subsection{Homogeneous Differentiator tuning procedure}
\label{sec:self-tuning_Diff}

The HD system is a differentiator for scalar signals (possibly affected by noise) whose state variables, under suitable assumptions, converge in \emph{finite time} to the corresponding derivatives of the underlying noise-free signal; in the ideal noiseless case the estimation error vanishes, whereas in the presence of noise 
the estimation error is guaranteed to be bounded, with bounds determined by the noise level and design parameters. A central parameter of HD is the gain $L>0$, which controls how the HD processes the data and sets the trade-off between faster convergence to the derivative estimates and noise reduction.
Here we describe only the aspects of HD that are needed to present the gain selection procedure, while the remaining details of its continuous-time and discrete-time formulations are reported in Supplementary Section S1.1.
The proposed procedure for inferring $L$ from the data, corresponding to the “Inference from data” block acting on the Differentiator block in Fig.~\ref{fig:methodology_pipeline}(b), is summarised in Algorithm~\ref{alg:hd_tuning}.

\subsubsection{Homogeneous Differentiators}

Let $y(t)=\bar{y}(t)+\eta(t)$ denote the measured scalar signal, where $\bar{y}(t)$ is given in \eqref{eq:generic_ode} and $\eta(t)$ models measurement noise.
An HD is a nonlinear system with discontinuous right-hand side that, given the noisy input $y(t)$ and a fixed differentiation order $n_d\in\mathbb{N}$, produces $\boldsymbol{z}(t)=[z_0(t),\ldots,z_{n_d}(t)]^\top$, whose components approximate the derivatives $\bar{y}^{(i)}(t)$, $i\in [n_d]_0$, in finite time.
The continuous-time dynamics of the HD can be written as 
\begin{equation}
  \dot{\boldsymbol{z}}(t)=\mathcal{D}_{n_d}\bigl(y(t),\boldsymbol{z}(t),L\bigr),
\label{eq:HomDiff_compact}
\end{equation}
where $L>0$ is the HD gain, which is a \emph{user-defined parameter} (see Supplementary Section S1.1).
Convergence of the HD estimates to the derivatives of the clean signal is ensured under appropriate assumptions regarding the differentiability of the noise-free signal $\bar{y}(t)$, whose $(n_d+1)$-th derivative must be upper bounded as $\bigl|\bar{y}^{(n_d+1)}(t)\bigr| \leq L$ for all $t$, and the essential boundedness of the noise $\eta(t)$. The precise assumptions are reported in Supplementary Section S1.1; see also \cite{levant2003higher,levant2020robust}.
Under such assumptions, the HD output $z_i(t)$ converges to the corresponding derivative $\bar{y}^{(i)}(t)$, with rigorous bounds on the estimation error $\varepsilon_i=|z_i-\bar{y}^{(i)}|$. 
Crucially, the bounds on $\left\{\varepsilon_i \right\}_{i=0}^{n_d}$ depend on $L$ (see Supplementary Section S1.1).
Intuitively, increasing $L$ makes $z_i$ converge more rapidly to $\bar{y}^{(i)}$, but also generates high-frequency oscillations around zero in $\varepsilon_i$,
which are 
induced by noise and by the discontinuities of HD's right-hand side; this behaviour is commonly referred to as \emph{chattering} \cite{levant2003higher}.
Conversely, decreasing $L$ reduces chattering at the cost of slower convergence and of larger error bounds on $\left\{\varepsilon_i \right\}_{i=0}^{n_d}$.
This convergence-chattering trade-off is central to HDs: the simulations in Fig.~\ref{fig:diff_autotuning} show that choosing naively the gain $L$ can yield derivative estimates that remain noticeably far from the actual derivatives when $L$ is too small, or are affected by severe chattering when $L$ is too large. Thus, a fully automatic and data-driven gain selection procedure, which avoids manual parameter tuning, is crucial for implementability of HD in applications.

Since realistic data is available as sampled time series, we employ a discretised HD, which processes sampled measurements $y_k=y(t_k)$, $k\in [K]_0$, with uniform sampling period $\Delta>0$ (for simplicity).
The HD is implemented by the forward-Euler scheme
\begin{equation}
  \boldsymbol{z}_{k+1}
  = \boldsymbol{z}_k
    + \Delta\mathcal{D}_{n_d}(y_k,\boldsymbol{z}_k,L)
    + \boldsymbol{T}_{n_d}(\boldsymbol{z}_k,\Delta),
\label{eq:Discrete_HD}
\end{equation}
where $\boldsymbol{T}_{n_d}$ contains correction terms that help preserve the accuracy of the continuous-time differentiator in \eqref{eq:HomDiff_compact} after discretisation (see Supplementary Section S1.1).
The total error, i.e. the discrepancy between the discrete-time HD output and the continuous-time derivatives they approximate, combines two effects: the first comes from the continuous-time HD itself and reflects the trade-off between fast convergence and chattering discussed above, while the second is introduced by sampling and discretisation and becomes smaller as the sampling period $\Delta$ decreases.
In all the performed simulations, we also use the low-chattering modification proposed in \cite{hanan2021low} with parameter $q=0$, which reduces chattering when the HD output is already close to the time derivatives of the noise-free signal, while preserving differentiation accuracy (see Supplementary Fig. S5 for examples of the effect of the low-chattering modification). The problem of gain selection is equally crucial for the performance of the discrete-time HD (as for the continuous-time HD). Importantly, in this work we focus on how to choose $L$ automatically from $\{y_k\}_{k=0}^K$ alone, without access to the noise-free signal $\{\bar{y}_k\}_{k=0}^K$.

\subsubsection{Residual-based inference from data of the HD gain}

\begin{figure}[ht!] 
\centering\includegraphics[width=1\linewidth]{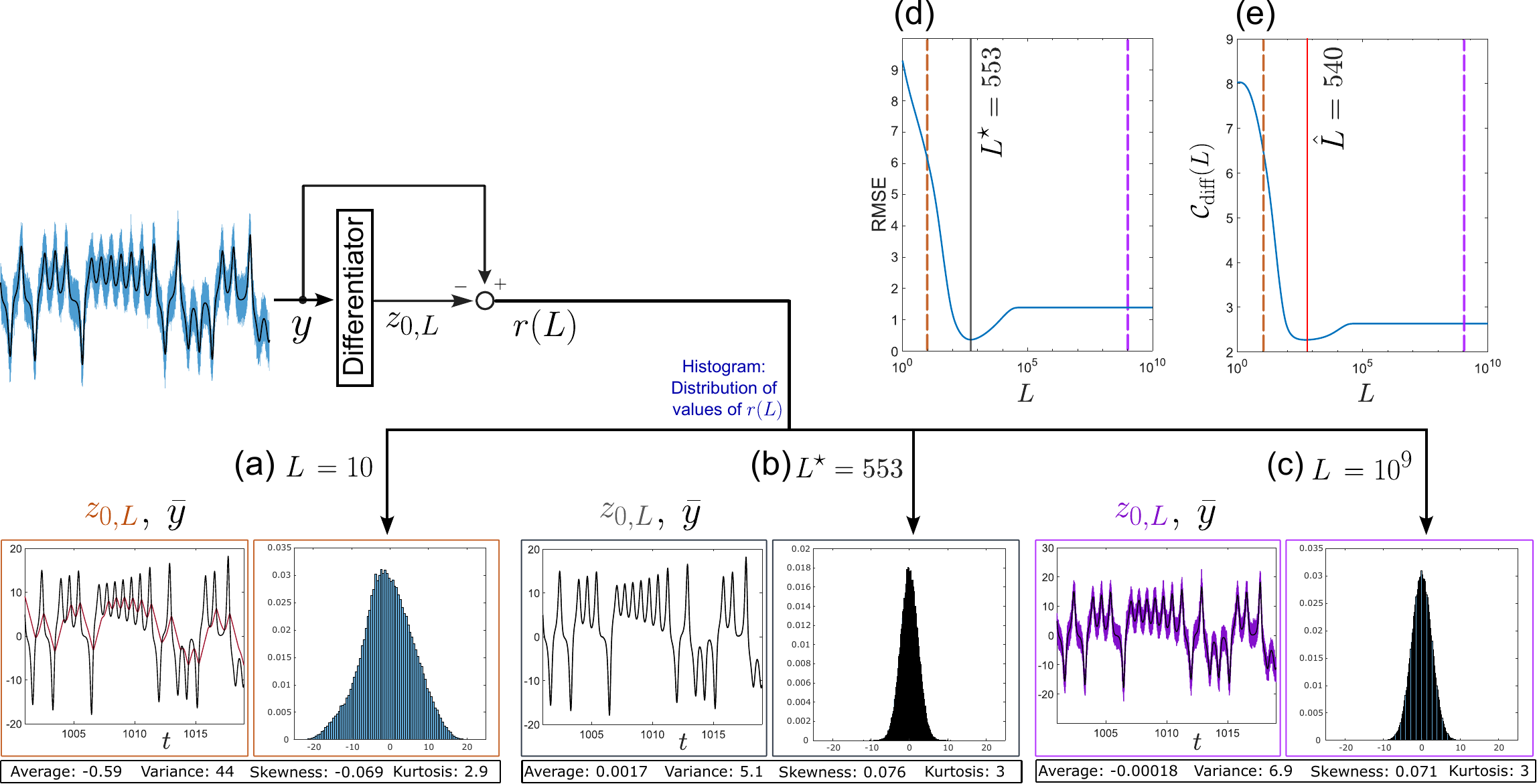}
\caption{\textbf{SHADED selects an HD gain close to the inaccessible RMSE-optimal value using a residual-based cost that requires no ground truth.}
Automatic inference of the HD gain $L$ for zeroth-order differentiation ($n_d=0$) of the Lorenz system output; see Supplementary Table S1 and Fig.~\ref{fig:methodology_pipeline} for details. In panels (a), (b), (c), the left plot shows the estimated $z_{0,L}$ (in colour), reconstructed from the noisy signal $y$, compared to the (inaccessible) noise-free signal $\bar y$ (in black), while the right plot shows a histogram with the distribution of values of the residual $r(L)=y-z_{0,L}$.
(a) With an underestimated gain $L=10 \ll L^\star=553$ (orange), the HD output $z_{0,L}$ does not faithfully track the dynamics of the underlying signal and the residual $r(L)=y-z_{0,L}$ retains a strong low-frequency component, leading to a significant increase in residual variance.
(b) With the RMSE-optimal gain $L^\star=553$ (gray), the HD output closely follows the noise-free signal $\bar{y}$ without introducing noticeable chattering.
(c) With an overestimated gain $L=10^9 \gg L^\star=553$ (purple), the HD output exhibits pronounced chattering, which is visible in the rapid oscillations around the noise-free signal and is consistent with an increase in residual variance.
(d) RMSE between $z_{0,L}$ and $\bar{y}$ over $L\in[10^0,10^{10}]$, for 1000 logarithmically equally spaced points; the gray vertical line corresponds the ideal gain $L^\star=553$ that minimises the RMSE, while the orange and purple dashed vertical lines respectively correspond to the underestimated $L=10$ and the overestimated $L=10^9$.
(e) Cost function $\mathcal{C}_{\mathrm{diff}}(L)$ in \eqref{eq:diff_cost} evaluated over the same interval of $L$ as in panel (d); the red vertical line corresponds to the gain $\hat{L}=540$ automatically selected by SHADED, which is close to $L^\star$: the example demonstrates that $\mathcal{C}_{\mathrm{diff}}(L)$ is a good proxy for the RMSE and minimising $\mathcal{C}_{\mathrm{diff}}(L)$ yields a gain comparable to that of the ideal, but inaccessible, RMSE-based choice.}
\label{fig:diff_autotuning}
\end{figure}

Let $\{y_k\}_{k=0}^K$ be the measured signal and $\{z_{i,L,k}\}_{k=0}^K$, $i\in [n_d]_0$, denote the $i$-th component of the HD output obtained from \eqref{eq:Discrete_HD}, which approximates the samples of the derivative $\bar{y}^{(i)}$ of the noise-free signal for a chosen gain $L$ and all $k\in[K]_0$. 
Define the differentiator residual as
\begin{equation}
  r_k(L) = y_k - z_{0,L,k}, \quad k\in [K]_0,
\label{eq:diff_residual_dt}
\end{equation}
where $z_{0,L,k}$ is the zeroth-order component of the HD output. 
The differentiator residual serves as a proxy for the noise-free tracking error $\left\{\bar{y}_k-z_{0,L,k} \right\}_{k=0}^K$ and quantifies the accuracy of tracking of the measured data by the HD output for a candidate gain $L$. The differentiator residual is the basis of the residual-based inference from data block shown in Fig.~\ref{fig:methodology_pipeline}(b). 
In fact, the sequence $\{r_k(L)\}_{k=0}^K$ reflects the same gain-dependent behaviour as the noise-free tracking error. When $L$ is too small, the residual $\{r_k(L)\}_{k=0}^K$ retains a persistent low-frequency component, which is contained in $y$ due to incomplete tracking of the component $\bar{y}$, since the HD output does not track the measured data closely enough. Conversely, when $L$ is too large, $\{z_{0,L,k}\}_{k=0}^K$ eventually tracks $\{\bar{y}_k\}_{k=0}^K$, but is accompanied by strong chattering, so the residual $\{r_k(L)\}_{k=0}^K$ is composed of high-frequency oscillations, introduced by the noise, which fluctuate around zero with an amplitude that increases with $L$ (see Fig.~\ref{fig:diff_autotuning} for the case of HD with $n_d=0$).
To quantify these effects, we compute the sample standard deviation of the residual:
\begin{equation}
  \sigma(L) = \sqrt{\frac{1}{K}\sum_{k=0}^{K}
  \Biggl( r_k(L) - \frac{1}{K+1}\sum_{j=0}^{K} r_j(L) \Biggr)^{\!2}}.
\label{eq:residual_statistics}
\end{equation}

In addition to the measurement noise, the residual contains a structured component, due to low-frequency tracking error for small $L$ and to high-frequency chattering for large $L$, that increases the sample standard deviation $\sigma(L)$.
For intermediate gains, by contrast, the HD output tracks the measured signal closely without introducing noticeable chattering, so the residual is dominated by the measurement noise and does not exhibit such additional structured components.
Hence, a large $\sigma(L)$ reveals either under- or over-tuning, and we define the differentiator cost function as the residual standard deviation,
\begin{equation}
  \mathcal{C}_{\mathrm{diff}}(L) = \sigma(L),
\label{eq:diff_cost}
\end{equation}
which has the same physical units as the signal and can be interpreted as a measure of residual variability.
In the theoretical models considered in Section~\ref{sec:results}, where the ground truth is accessible for comparison, minimising $\mathcal{C}_{\mathrm{diff}}(L)$ consistently returns gains that are closer to the RMSE-optimal value $L^\star$ than those obtained by alternative residual-based costs that also incorporate the sample mean of the residual (see Supplementary Figs. S1-S4).
For this reason, only the standard-deviation-based cost \eqref{eq:diff_cost} is used in what follows.
Therefore, the automatically inferred gain is defined as
\begin{equation}
  \hat{L} = \operatorname*{arg\,min}_{L\in\Lambda} \mathcal{C}_{\mathrm{diff}}(L),
  \quad
  \Lambda = [L_{\min},L_{\max}] \subset (0,+\infty).
\label{eq:L_hat_def}
\end{equation}
Here, $\Lambda$ defines the admissible range of gains explored by our algorithm, with $L_{\max}$ determined automatically from the measured data, so that $\hat{L}$ is obtained directly from $\{y_k\}_{k=0}^K$.
In the HD framework, the HD gain $L$ serves as a Lipschitz bound for the $n_d$-th derivative of the noise-free signal $\bar{y}$, so the $(n_d+1)$-th derivative of $\bar{y}$ is bounded in magnitude by the constant $L$ \cite{levant2003higher,levant2020robust}; see also Supplementary Section S1.1.
The lower bound $L_{\min}>0$ excludes the degenerate case $L=0$. In fact, $L=0$ can only hold if $\bar{y}^{(n_d+1)}\equiv 0$, meaning that $\bar{y}$ is a polynomial of degree at most $n_d$, and only two possibilities arise.
Either $\bar{y}$ is constant, in which case the measurements do not vary at all and carry no information about the underlying dynamics, so that no reconstruction is possible in the first place; or $\bar{y}$ is a nonconstant polynomial and therefore is unbounded.
The latter case is incompatible with the problem considered here: we aim to reconstruct a compact attractor of \eqref{eq:generic_ode}, up to a continuous change of coordinates, so that by assumption $\boldsymbol{x}(t)$ evolves in a bounded neighbourhood of that attractor, and since $h$ is smooth, $\bar{y}(t)=h(\boldsymbol{x}(t))$ is necessarily bounded.
Consistently, in all the examples we studied, the empirically relevant gains satisfy $L>1$, confirming that gains close to zero do not arise in practice (see Supplementary Figs. S1-S4); we thus set $L_{\min}=1$.
We further estimate $L_{\max}$  by computing the $(n_d+1)$-th order finite difference of $\{y_k\}_{k=0}^K$, using a central difference at interior points and forward or backward differences at the boundaries, according to the samples available. Denoting the finite difference operator applied in the neighbourhood of a point $y_k$ as $\mathfrak{D}^{(n_d+1)}y_{k}$, we set
\begin{equation}
  L_{\max} = \max\{10\cdot\max_{k \in [K]_0} 
  \bigl|\mathfrak{D}^{(n_d+1)} y_k\bigr|,1\}.
\label{eq:L_max}
\end{equation}
The quantity $\max_{k \in [K]_0}|\mathfrak{D}^{(n_d+1)}y_k|$ serves as a data-driven proxy for the bound on the magnitude of the $(n_d+1)$-th derivative of the noise-free signal $\bar{y}$; since $\bar{y}$ is expected to be smooth, the finite-difference estimate captures also the variation introduced by (non-smooth) noise and thus tends to overestimate the magnitude of the $(n_d+1)$-th derivative of $\bar{y}$. The factor $10$ is further introduced to provide a wider margin for the estimation of $L_{\max}$, ensuring it is indeed an upper bound.
Note that the generated finite differences are \emph{not} used to approximate the derivatives of the noise-free signal, but are used only to obtain the upper bound $L_{\max}$ on $L$.

Since the optimal value of $L$ is not known a priori and may span several orders of magnitude in different applications (see Fig.~\ref{fig:diff_autotuning}(e)), to improve numerical robustness, we minimise \eqref{eq:diff_cost} in the base-10 logarithmic domain.
After computing $L_{\max}$ from \eqref{eq:L_max}, we introduce the normalised optimisation variable $\xi = \log_{10}(L)/\log_{10}(L_{\max})\in[0,1]$, so that $L = L_{\max}^{\xi}$.
This parametrisation turns the search into a logarithmic one and restricts attention to gains $L\in[1,L_{\max}]$.  
This is consistent with the role of $L$ as a Lipschitz constant and with the fact that, for the theoretical models considered in Section~\ref{sec:results_models} (where the RMSE-optimal gain $L^\star$ can be assessed directly), the practically relevant HD gains never fall below $1$. 
We then apply simulated annealing \cite{kirkpatrick1983optimization}, a stochastic optimisation method that occasionally accepts worse moves to reduce the risk of local minima, to the scalar function $\xi\mapsto\mathcal{C}_{\mathrm{diff}}(L_{\max}^{\xi})$ on the unit interval.
The search is initialised at $\xi=1$ and, throughout the case studies in Section~\ref{sec:results}, we use the following default stopping criterion: the optimisation terminates when the variation of $\mathcal{C}_{\mathrm{diff}}$ between two successive candidate values selected by the algorithm falls below $10^{-4}$ or when $150$ iterations are reached, whichever occurs first (see Supplementary Fig. S10 for the performance study with these settings, where the inferred gains are concentrated around values that are approximately the same across independent simulated-annealing runs).
The stopping criterion can be adjusted if a finer or coarser search is desired.
The resulting normalised optimum $\hat{\xi}\in[0,1]$ is then mapped back to the original scale, finally yielding $\hat{L} = L_{\max}^{\hat{\xi}}$.

\begin{algorithm}[t]
\caption{Automatic inference of the \textbf{HD gain $L$} from data}
\label{alg:hd_tuning}
\begin{algorithmic}[1]
\State Compute $L_{\max}$ via \eqref{eq:L_max} from the $(n_d+1)$-th order central finite difference of the noisy signal $\{y_k\}_{k=0}^{K}$.
\State Introduce the normalised variable $\xi = \log_{10}(L)/\log_{10}(L_{\max}) \in [0,1]$.
\State Determine $\hat{\xi} = \operatorname*{arg\,min}_{\xi \in [0,1]} \mathcal{C}_{\mathrm{diff}}\!\left(L_{\max}^{\xi}\right)$, where $\mathcal{C}_{\mathrm{diff}}$ is defined in \eqref{eq:diff_cost}, using simulated annealing.
\State Recover the gain inferred from data as $\hat{L} = L_{\max}^{\,\hat{\xi}}$.
\end{algorithmic}
\end{algorithm}

\subsection{Savitzky-Golay filter window size inference from data}
\label{sec:self-tuning_SG}

The output of the HD with the selected $\hat{L}$ may still contain chattering, due to the HD dynamics. 
We thus process the HD output with an SG filter (see \cite{sutulovic2025efficient} for a discussion of the improvement in attractor reconstruction thanks to the SG filter) 
whose window size is chosen so as to maximally reduce chattering while preserving the relevant features of the noise-free signal, using the procedure summarised in Algorithm~\ref{alg:sg_tuning} and illustrated in Fig.~\ref{fig:SG_autotuning}. In Fig.~\ref{fig:methodology_pipeline}(b), the HD output is processed by the smoothing SG block, with SG window size automatically selected (“Inference from data” block acting on the SG block) using our proposed procedure.

\subsubsection{Savitzky-Golay filter}

We first briefly recall the classical SG filter theory. 
The classical SG filter  computes at each index $k$ the value of a polynomial of fixed degree $d\in\mathbb{N}$ fitted in a least-squares sense over a sliding window centred at the $k$th signal sample \cite{savitzky1964smoothing}.
Throughout this work we set $d = 2$, so that the SG filter uses a quadratic polynomial fit, in line with the observation that low-degree polynomials are adequate for smoothing with the SG filter \cite{kennedy2020improving}. Indeed, 
frequency-domain analyses show that low-order SG filters already preserve the signal’s low-frequency content with little attenuation, and are therefore usually sufficient for capturing the essential information of the signal, while higher polynomial orders can increase sensitivity to noise \cite{schafer2011savitzky,luo2005properties}. 
Thus, in this manuscript, the automatic procedure focuses on selecting the SG filter window size, while $d=2$ is kept fixed.

Let $\{x_k\}_{k=0}^{K}$ denote the signal to be filtered, which may be any entry of the HD output obtained from \eqref{eq:Discrete_HD} with the tuned gain $\hat{L}$ selected as in Section~\ref{sec:self-tuning_Diff}.
This notation is introduced to keep the SG tuning procedure general, since the same filtering rule is applied to any component of the HD output at any step of the computational methodology; see Fig.~\ref{fig:methodology_pipeline}(c).
For a given odd window size $w\in\mathbb{N}$, the SG filter operates on a symmetric neighbourhood of each data point: setting $m=(w-1)/2$, we define the admissible index set of the time series $\mathcal{I}_w=\{m,m+1,\ldots,K-m\}$. The resulting SG filter will be applied only to data points $x_k$ with $k\in \mathcal{I}_w$, with $w$ regulating the number of samples around $x_k$ that are used in the local polynomial fit.
For each $k\in\mathcal{I}_w$, the SG filter acts as a linear time-invariant finite-impulse-response operator whose frequency response depends on the polynomial degree $d$ and on the window size $w$:
\begin{equation}
  \tilde{x}_k(w)=\sum_{j=-m}^{m} a_j^{(w)}\,x_{k+j},
  \quad k\in\mathcal{I}_w,
\label{eq:sg_filtering}
\end{equation}
where the coefficients $\boldsymbol{a}^{(w)}=[a_{-m}^{(w)},\ldots,a_m^{(w)}]^\top$ depend on the chosen polynomial degree $d$ and are obtained from the corresponding local least-squares problem on the symmetric window of $w$ samples centred at $x_k$  \cite{schafer2011savitzky}. 
Although SG filters are not designed as sharp low-pass filters, they exhibit moderate stopband attenuation and approximately low-pass behaviour \cite{luo2005properties,schmid2022and}: for fixed $d$,
small $w$ keeps the neighbourhood narrow and preserves higher-frequency signal variations, at the cost of limited chattering suppression, whereas large $w$ averages over a broader neighbourhood, which attenuates high-frequency signal fluctuations more strongly, but may distort the underlying signal by filtering out crucial high-frequency signal components that contain information about the noise-free signal derivatives and, as a result, about the attractor to be reconstructed.

To reduce boundary artifacts introduced by SG filtering \cite{schmid2022and} and to exclude the initial portion of the record where the HD output may still be affected by its finite-time transient \cite{levant2003higher}, we discard samples symmetrically from both the beginning and the end of each time series. 
Specifically, given an estimate $\tau_{\mathrm{tr}}$ of the HD transient duration, we retain only the subsegment $[t_0+\tau_{\mathrm{tr}},\, t_K-\tau_{\mathrm{tr}}]$, so that all retained samples lie strictly inside the record and are processed with fully centred SG filter windows.
When no problem-specific value is available, we set $\tau_{\mathrm{tr}}$ to be $5\%$ of the total record length, as a simple heuristic. 
This symmetric trimming removes both the edge effects of the SG filter and the HD transient without introducing an additional tuning parameter; studying the optimal choice of $\tau_{\mathrm{tr}}$ is left for future work.

\subsubsection{Cost-based window selection}

The procedure for the automatic selection of the SG window size $w$ first identifies a candidate window by minimising a cost function that balances chattering suppression and preservation of the low-frequency content of the signal, under a frequency-separation assumption: 
specifically, we assume that the chattering component is concentrated at high frequencies, whereas the informative components of the signal lie in the low-frequency range, and we seek the smallest window $w$ that only removes the high-frequency chattering. 
We define an SG cost $\mathcal{C}_{\mathrm{SG}}(w)$, shown in \eqref{eq:sg_cost} below, that trades off two competing effects of increasing the window size. 
As $w$ grows, the SG filter removes more of the high-frequency chattering, which is desirable; however, once $w$ becomes too large, it also starts attenuating the low-frequency content of the signal, which is undesirable. 
The cost $\mathcal{C}_{\mathrm{SG}}(w)$ is therefore constructed to decrease while the increase in $w$ mainly suppresses chattering, and to increase once further growth of $w$ begins to remove low-frequency signal content.

An optional adjustment, applied after the cost-based selection has produced a candidate window $w'$ and aimed at further reducing the risk of oversmoothing, can subsequently adjust $w'$ based on spectral properties of the SG residual, defined as the difference between the HD output and its SG-filtered version.
Starting from $w'$, we decrease the window size while monitoring the dominant frequency of the residual, and we select the smallest window that still preserves such dominant frequency in the residual, and hence removes it from the SG-filtered signal.

Both the first selection and the optional adjustment operate directly on the HD output and are repeated at each staircase level in Fig.~\ref{fig:methodology_pipeline}(c). Thus, all SG window sizes are obtained from the data. The SG cost function $\mathcal{C}_{\mathrm{SG}}(w)$ is built by combining two dimensionless quantities. 

The first quantity measures chattering suppression. Given $\{y_k\}_{k\in\mathcal{I}_w}$, a derivative order $i$ and an HD gain $L$, we denote $x_k^{(L)} = z_{i,L,k}$, $k \in \mathcal{I}_w$, where $\{z_{i,L,k}\}_{k\in\mathcal{I}_w}$ is the $i$-th component produced by \eqref{eq:Discrete_HD} from $\{y_k\}_{k\in\mathcal{I}_w}$ with gain $L$. Recalling the estimated HD gain $\hat{L}$, 
we compare two HD outputs obtained from the  input $\{y_k\}_{k\in\mathcal{I}_w}$ and two nearby gains $\hat{L}$ and $(1+\epsilon)\hat{L}$, $\epsilon>0$, that is, $\{x_k^{(\hat{L})}\}_{k\in\mathcal{I}_w}$ and $\{x_k^{((1+\epsilon)\hat{L})}\}_{k\in\mathcal{I}_w}$, respectively. 
For each window size $w$, we then apply the SG filter \eqref{eq:sg_filtering} to these HD outputs, obtaining the SG-filtered sequences $\{\tilde{x}_k^{(\hat{L})}(w)\}_{k\in\mathcal{I}_w}$ and $\{\tilde{x}_k^{((1+\epsilon)\hat{L})}(w)\}_{k\in\mathcal{I}_w}$. 
We then define the chattering-sensitive sequence $\{c_k(w)\}_{k\in\mathcal{I}_w}$ as the pointwise difference between these two SG-filtered signals, namely $c_k(w) = \tilde{x}_k^{(\hat{L})}(w) - \tilde{x}_k^{((1+\epsilon)\hat{L})}(w)$, $k\in\mathcal{I}_w$. 
Intuitively, $c_k(w)$ captures how much the SG-filtered estimate of the $i$-th derivative changes when the HD gain is perturbed from $\hat{L}$ to $(1+\epsilon)\hat{L}$ while the SG window $w$ is kept fixed. 
In view of the assumption that was made about the frequency content separation between the clean signal and the noise, the two SG-filtered signals share the same low-frequency content and differ mainly in their high-frequency chattering component induced by the different HD gains. Hence, their difference $\{c_k(w)\}_{k\in\mathcal{I}_w}$ isolates the chattering contribution that depends on the gain and
serves as a direct, window-dependent measure of the chattering present in the SG-filtered HD output.
This interpretation is consistent with the differentiator cost function in \eqref{eq:diff_cost}, which increases near the minimiser $\hat{L}$ primarily due to the residual variance term, indicating that small gain perturbations to the right of $\hat{L}$ modify mostly the high-frequency oscillatory content, associated with chattering, while leaving the low-frequency component practically unchanged (see Supplementary Section S1.2 for additional examples and further discussion).
We select $\epsilon=0.05$, although varying $\epsilon$ in multiple test cases produced no effect in the obtained window sizes (see Supplementary Fig. S7).
We quantify the variability of $\{c_k\}_{k\in\mathcal{I}_w}$ through its discrete total variation, $\operatorname{TV}_{\mathrm{chat}}(w) =\frac{1}{|\mathcal{I}_w|-1}\sum_{k\in\mathcal{I}_w\setminus \left\{\max \mathcal{I}_w \right\}}|c_{k+1}-c_k|$, which measures the oscillatory content of $\{c_k\}_{k\in\mathcal{I}_w}$: larger values correspond to larger sample-to-sample changes and thus to stronger high-frequency components, whence $\operatorname{TV}_{\mathrm{chat}}(w)$ can be used as a proxy for the spectral content of the chattering component.
The normalisation factor $|\mathcal{I}_w|-1$ divides the cumulative absolute difference by the number of adjacent pairs in the window, so that $\operatorname{TV}_{\mathrm{chat}}(w)$ represents an average sample-to-sample change rather than a quantity that grows automatically with the length of $\mathcal{I}_w$.  
This allows us to compare different window sizes $w$ on a common scale, focusing on how strongly the sequence $\{c_k(w)\}_{k\in\mathcal{I}_w}$ oscillates instead of simply reflecting the number of samples over which it is computed.
Since SG filters act approximately as low-pass operators whose window size predominantly affects how much high-frequency content is attenuated \cite{luo2005properties,schmid2022and}, $\operatorname{TV}_{\mathrm{chat}}(w)$ is expected to decrease as the SG window size $w$ increases and chattering is removed, and thereby quantifies the effectiveness of chattering suppression (see Supplementary Fig. S6).

The second quantity penalises excessive smoothing and balances the tendency of $\operatorname{TV}_{\mathrm{chat}}(w)$ to favour larger SG filter windows, which may remove relevant signal content. 
For each window size $w$, we apply the SG filter \eqref{eq:sg_filtering} to the HD output component of interest $\{x_k^{(\hat{L})}\}_{k\in\mathcal{I}_w} = \{ z_{i,\hat{L},k} \}_{k\in\mathcal{I}_w}$, and obtain the filtered sequence $\{\tilde{x}_k^{(\hat{L})}(w)\}_{k\in\mathcal{I}_w}= \{ \tilde z_{i,\hat{L},k} \}_{k\in\mathcal{I}_w}$.
When $w$ becomes very large, the SG filter suppresses not only chattering but also part of the lower-frequency content of the signal, and the sequence $\{\tilde{x}_k^{(\hat{L})}(w)\}_{k\in\mathcal{I}_w}$ becomes nearly constant or very slowly varying.  
In this situation, the variance of $\{\tilde{x}_k^{(\hat{L})}(w)\}_{k\in\mathcal{I}_w}$ decreases, indicating loss of low-frequency content in addition to suppression of high-frequency fluctuations.
We therefore set
$  D_{\mathrm{sig}}(w)=\frac{1}{\operatorname{Var}(\tilde{x}^{(\hat{L})}(w))},$
where $\operatorname{Var}(\tilde{x}^{(\hat{L})}(w))$ is the sample variance of $\{\tilde{x}_k^{(\hat{L})}(w)\}_{k\in\mathcal{I}_w}$. 
A smaller variance indicates stronger smoothing, whence $D_{\mathrm{sig}}(w)$ increases with the window size, when the SG filter progressively suppresses not only chattering but also the frequency components of the chattering-free signal. 
Therefore, $D_{\mathrm{sig}}(w)$ acts as a penalty for window sizes that cause excessive signal smoothing (see Supplementary Fig. S6).

Let $w_{\min}$ be the smallest admissible SG filter window, which by default is the first odd integer greater than $d$, and let $w_{\max}$ be an upper bound on the window size, which we set to $w_{\max}=20001$ to maintain the computational efficiency of the SG filtering in the examples considered in Section~\ref{sec:results}.
The sensitivity analysis in Supplementary Fig. S9 shows that the inferred windows do not depend very strongly on this choice: when $w_{\max}$ is varied between $10001$ and $20001$, the selected window size changes by at most 44\% relative to its value at the default $w_{\max}$. 
When additional information about the time scales of the system is available, $w_{\max}$ can 
be chosen manually so as to match the time scale over which the signal varies in a meaningful way.
Because $\operatorname{TV}_{\mathrm{chat}}(w)$ and $D_{\mathrm{sig}}(w)$ may differ by several orders of magnitude, we normalise them before combining them into the designed cost function.
As mentioned before, as $w$ increases, the SG filter removes more high-frequency content, so $\operatorname{TV}_{\mathrm{chat}}(w)$ is expected to be largest for the smallest admissible window $w_{\min}$, whereas $D_{\mathrm{sig}}(w)$ is expected to be largest for the largest admissible window $w_{\max}$
(see Supplementary Fig. S6).
We therefore define the SG cost function as
\begin{equation}
  \mathcal{C}_{\mathrm{SG}}(w)
  =(1-\lambda)\frac{\operatorname{TV}_{\mathrm{chat}}(w)}{\operatorname{TV}_{\mathrm{chat}}(w_{\min})}
  +\lambda\frac{D_{\mathrm{sig}}(w)}{D_{\mathrm{sig}}(w_{\max})},
  \quad \lambda\in(0,1),
\label{eq:sg_cost}
\end{equation}
where smaller values of $\lambda$ place more emphasis on chattering suppression, captured by $\operatorname{TV}_{\mathrm{chat}}$, whereas larger values of $\lambda$ favour preservation of the signal variance, captured by $D_{\mathrm{sig}}$. 
To obtain a meaningful trade-off between these two objectives, $\lambda$ should be chosen away from the extremes $0$ and $1$.
In all examples reported in this paper, we set $\lambda=0.5$ by default, giving equal weight to both objectives and we verify empirically that the selected window sizes exhibit relatively modest changes when $\lambda$ is varied moderately around this value for the examples considered in Section~\ref{sec:results_models} (see Supplementary Fig. S8).
By definition, smaller values of $\mathcal{C}_{\mathrm{SG}}(w)$ correspond to window sizes that are expected to reduce the high-frequency content associated with chattering while keeping the variance of the filtered signal within a range that preserves the low-frequency content of the original signal.

Rather than evaluating $\mathcal{C}_{\mathrm{SG}}(w)$ on a fixed discrete grid, we parametrise the window by its associated time span $T=w\Delta\in[T_{\min}, \, T_{\max}]$, where $T_{\min}=w_{\min}\Delta$ and $T_{\max}=w_{\max}\Delta$, and treat $T$ as a continuous optimisation variable. 
We minimise $\mathcal{C}_{\mathrm{SG}}(\frac{T}{\Delta})$ over $T\in[T_{\min}, \, T_{\max}]$ using simulated annealing \cite{kirkpatrick1983optimization} with minimum cost variation of $10^{-4}$ and $50$ maximum iterations (see Supplementary Fig. S10 for the performance study with these settings, where the cost-based SG windows are concentrated around values that are approximately the same across independent simulated-annealing runs). 
Let $T'$ be the obtained minimiser.
We convert it into the corresponding odd SG window length by setting
\begin{equation}
  w'=\bar{w}(T')
   =2\left\lfloor\frac{T'}{2\Delta}\right\rfloor+1,
\label{eq:w_prime_def}
\end{equation}
which is the candidate SG filter window produced by the cost-based stage. 
At this point, one may use $w'$ directly as the SG window, relying solely on the cost-based optimisation to balance chattering suppression and low-frequency content preservation.  
Alternatively, $w'$ can be further adjusted by exploiting a persistence criterion on the dominant frequency of the SG residual, which monitors how the main spectral peak of the residual (associated with chattering) evolves as the window size is decreased and retains only windows for which this peak remains at the same frequency. 
This second option can be particularly useful when the frequency-separation assumption is not strictly valid, for instance when the signal contains relevant middle-frequency components that lie between the low-frequency informative content and the high-frequency chattering.  
In such cases, a window chosen only on the basis of the cost may be large enough to attenuate not only chattering but also part of these intermediate-frequency components, because the cost relies only on a global measure of signal variance and does not explicitly distinguish between different frequency bands in the signals involved.
As a result, it cannot tell whether the loss of variance is due to suppression of chattering at high frequencies or to unwanted attenuation of informative components at intermediate frequencies.
In this case we prioritise the preservation of such intermediate-frequency content at the cost of preserving additional chattering-induced content.

The usefulness of this adjustment is visible in the simulations of Section~\ref{sec:results}.
For the theoretical models of Section~\ref{sec:results_models}, where the frequency-separation assumption is largely satisfied, the cost-based window $T'$ is already close to the RMSE-optimal window $T^\star$ and the adjustment changes the reconstruction only marginally (see Fig.~\ref{fig:SG_autotuning} and Supplementary Figs. S13-S14 and S15-S16).
The benefit becomes evident in the examples where the chattering term of the cost remains large over a wide range of windows.
In the transistor-based chaotic circuits of Section~\ref{sec:results_LTspice} and in the cardiac recordings of Section~\ref{sec:results_data}, the cost alone tends to select relatively large windows that oversmooth the attractor, whereas the persistence-based adjustment selects a smaller window and yields an attractor whose derivative coordinates retain a substantially wider range and in which recurrent loops and nearby trajectory passages remain more clearly separated rather than merging into a densely filled region (compare Figs. S18-S19 with Figs. S20-S21, and Figs. S22-S24 with S25-S27, in the Supplementary Material).
For this reason the adjustment is an optional step in Algorithm~\ref{alg:sg_tuning}: it can be disabled when the frequency-separation assumption is expected to hold, and enabled when intermediate-frequency content is likely to be present, as is typically the case for real-life data, where intrinsic dynamics on multiple time scales and non-ideal noise generate spectral content between the low-frequency informative band and the high-frequency chattering.

\subsubsection{Persistent dominant-frequency tracking}
\label{sec:freq_persistence}
\begin{figure}[ht!] 
\centering\includegraphics[width=0.98\linewidth]{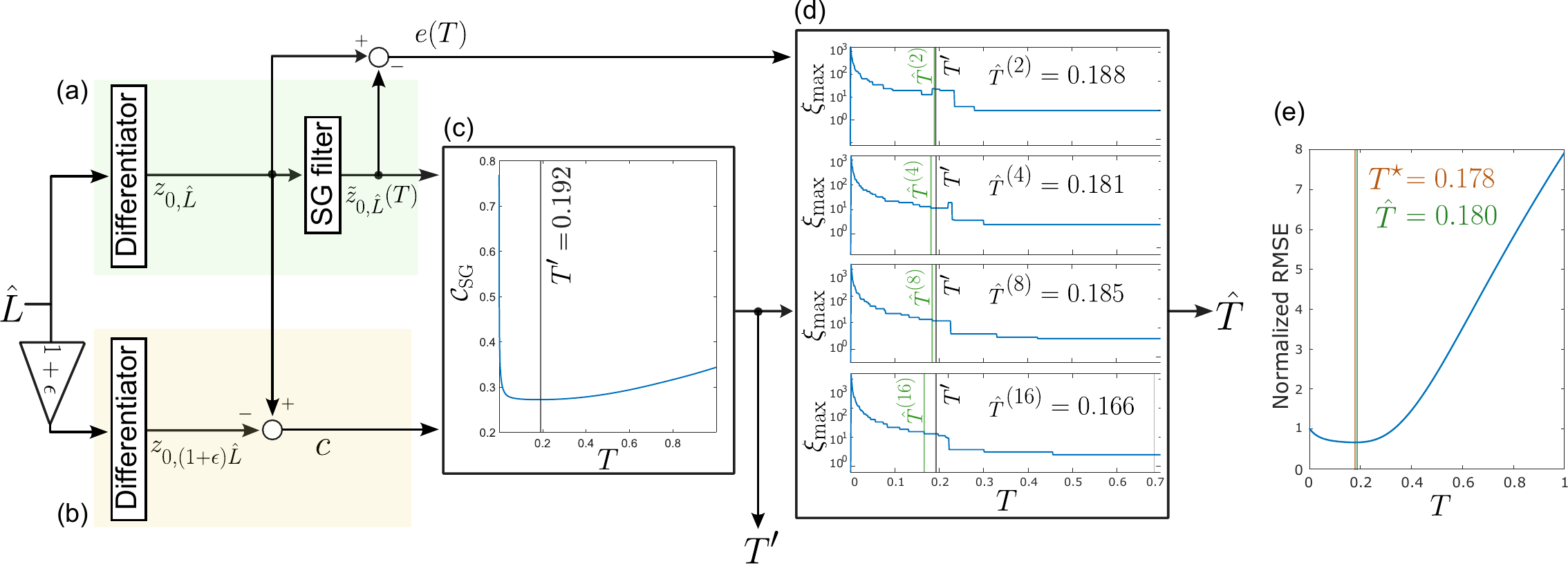}
\caption{\textbf{SHADED selects an SG window size close to the inaccessible RMSE-optimal value by combining signal-preservation with chattering suppression, and possibly with frequency persistence.}
Automatic inference of the SG filter window size $w$ with polynomial order $d=2$ for the HD zeroth-order derivative estimate with gain $\hat{L}$ and $n_d=0$ (as computed in Fig.~\ref{fig:diff_autotuning}) for the Lorenz system output $\bar{y}$; see Supplementary Table S1 and Fig.~\ref{fig:methodology_pipeline} for details. 
(a) Computation of the signal $\tilde z_{0,\hat{L}}$ (green) to assess
the component $D_{\mathrm{sig}}(w)$ of the SG cost $\mathcal{C}_{\mathrm{SG}}$.
(b) Computation of the chattering-sensitive signal $c$ (yellow) to assess
the component $\operatorname{TV}_{\mathrm{chat}}$ of the SG cost $\mathcal{C}_{\mathrm{SG}}$.
(c) SG cost $\mathcal{C}_{\mathrm{SG}}(w)=\mathcal{C}_{\mathrm{SG}}(\frac{T}{\Delta})$ defined in \eqref{eq:sg_cost} as a function of the SG filter window time span $T$ expressed in time units (for 1000 equally-spaced points), with $\Delta = 10^{-4}$; the vertical line corresponds to its minimiser $T'$. The selected window size can either be taken directly as $w'$, obtained from $T'$ through the formula \eqref{eq:w_prime_def}, or further adjusted as follows.
(d) Frequency $\xi_{\max}^{(N_{\mathrm{seg}})}(T)$ of the residual $e(T)$ in \eqref{eq:sg_residual} that maximises the PSD estimated with the Welch method using $50\%$ overlap and $N_{\mathrm{seg}}\in\{2,4,8,16\}$ segments. 
For each $N_{\mathrm{seg}}$, the window time span $T$ is decreased, starting from $T'$, until $\xi_{\max}(T)$ changes, and the smallest admissible window prior to the change is saved as $\hat{T}^{(N_{\mathrm{seg}})}$. 
The final window time span $\hat{T}$ is the average of $\hat{T}^{(j)},\ j=2,4,8,16$. The corresponding window size $\hat w$ is obtained by rounding to the nearest admissible odd integer via \eqref{eq:w_prime_def}.
(e) RMSE of the SG-filtered output $\tilde z_{0,\hat{L}}$ with respect to the noise-free signal $\bar{y}$, normalised by the RMSE of the corresponding HD output $z_{0,\hat{L}}$ without SG filtering, to quantify the improvement due to SG filtering.
The estimated $\hat{T}$ value is close to the ideal, but inaccessible, window time span $T^\star$ that minimises the RMSE.}
\label{fig:SG_autotuning}
\end{figure}

We describe here the optional adjustment that can be applied after the SG window time span $T'$ has been selected by minimising the cost $\mathcal{C}_{\mathrm{SG}}(\frac{T}{\Delta})$ in \eqref{eq:sg_cost}.
Given any SG filter window time span $T$, we define the SG residual as
\begin{equation}
  e_k(T)=x_k^{(\hat L)}-\tilde{x}_k^{(\hat L)}(T), \quad k\in\mathcal{I}_w.
\label{eq:sg_residual}
\end{equation}
Once $T$ is sufficiently large for the SG filter to suppress the chattering oscillations, the residual $\{e_k(T)\}_{ k\in\mathcal{I}_w}$ is largely composed of the high-frequency chattering fluctuations, as $\tilde{x}_k^{(\hat L)}(T)$ removes from $x_k^{(\hat L)}$ the low frequency signal components associated with the desired chattering-free signal. 
Therefore, for $T'$, we consider $e_k(T')$, $k\in \mathcal{I}_w$, and identify a pronounced peak in its power spectral density (PSD), corresponding to the frequency where the chattering energy is concentrated. This peak is observed consistently across neighbouring values of $T$ for which the SG residual retains predominantly the chattering component; see Fig.~\ref{fig:SG_autotuning}(d). Thus, we can decrease $T'$, while preserving the dominant peak in the residual PSD at the peak frequency identified for $T'$. As a result, decreasing $T$ reduces oversmoothing, while retaining the chattering reduction performed by the SG filter. Indeed, when $T$ becomes too small to remove the majority of the chattering, the PSD peak shifts, indicating that the filter is no longer removing the same oscillatory component.
We use this persistent residual frequency to adjust $T'$ and choose the smallest window that still removes chattering while preserving the lower-frequency content needed for faithful derivative estimation and attractor reconstruction.

Welch's method \cite{welch1967use} provides PSD estimates that are robust to noise and finite-length effects, and is a convenient tool for implementing the aforementioned idea.
We estimate the PSD by averaging periodograms (PSD estimates obtained from the magnitude-squared Fourier transform) computed on overlapping segments of $\{e_k(T')\}_{k\in\mathcal{I}_w}$, using $50\%$ overlap and several segment counts $\mathcal{N}_{\mathrm{seg}}=\{2,4,8,16\}$, which control the trade-off between spectral resolution and estimation reliability: fewer and longer segments preserve finer frequency detail, whereas more and shorter segments sacrifice frequency detail and yield an estimate that is less sensitive to the particular choice of segment boundaries. 
By checking several configurations, we reduce the risk that the selected window depends too strongly on a particular PSD estimate; instead, we retain a window only if the same dominant residual frequency can be identified consistently across segment counts, which makes the adjustment of $T'$ more robust. 

For each $N_{\mathrm{seg}}\in\mathcal{N}_{\mathrm{seg}}$, the same procedure is applied independently, yielding one tuned window per configuration.
We denote by $S_e(\xi,T)$ the PSD estimate at angular frequency $\xi\in[0,\xi_\nu]$, where $\xi_\nu=\pi/\Delta$ is the Nyquist angular frequency, and define
\begin{equation}
  \xi_{\max}(T)=\arg\max_{\xi\in[0, \, \xi_\nu]}S_e(\xi,T).
\label{eq:sg_residual_peak}
\end{equation}
Given the previously obtained $T'$, $\xi_{\max}(T')$ empirically exhibits a dominant peak at a characteristic chattering frequency $\xi'$, due to the presence of the chattering component in the residual.
As the window decreases from $T'$, $\xi_{\max}(T)$ remains close to $\xi'$ over a range of window sizes for which the residual is still mainly governed by the chattering component.
We therefore decrease the window size from $T'$ in steps of a user-specified decrement (by default $6$ samples per step) and, for each Welch configuration, select the smallest admissible window for which the dominant residual frequency remains equal to $\xi'$ (up to machine precision).
This procedure yields, for each $N_{\mathrm{seg}}\in\mathcal{N}_{\mathrm{seg}}$, a tuned window $\hat{T}^{(N_{\mathrm{seg}})}$ that is as short as possible, while still removing the same chattering band as at $T'$, and thus avoiding oversmoothing.
The final SG window time span $\hat T$ is obtained by averaging the windows selected for the different Welch configurations. The corresponding SG window size $\hat w$ is obtained by rounding the result to the nearest admissible odd integer through formula \eqref{eq:w_prime_def}, which makes the selection less sensitive to the particular choice of PSD estimation parameters.

Both the set $\mathcal{N}_{\mathrm{seg}}$ and the overlap percentage can be modified by the user when different spectral resolutions are desired.
When the record is long and the residual spectrum is noisy, increasing the number of segments (\textit{e.g.}, to $\{4,8,16,32\}$) improves estimation reliability at the cost of coarser frequency resolution, whereas when finer frequency detail is needed and the record is sufficiently long, reducing the number of segments (\textit{e.g.}, to $\{2,4,8\}$) yields sharper spectral features, in line with the classical trade-offs of Welch's method \cite{welch1967use}.
The overlap percentage can be increased (\textit{e.g.}, to $75\%$) to further reduce the variability of the PSD estimate from one segment configuration to another when long records are available, or decreased (\textit{e.g.}, to $25\%$) to reduce computational cost for very long time series.

\begin{algorithm}[t]
\caption{Automatic inference of the \textbf{SG window size} $w$ from data}
\label{alg:sg_tuning}
\begin{algorithmic}[1]
\State Given the sampling period $\Delta$, fix the HD gain to $\hat{L}$ as in Section~\ref{sec:self-tuning_Diff} and select the HD output to be processed. Set $R=1$ if residual-based adjustment is enabled, $R=0$ otherwise.
\State Choose the admissible SG window range $[w_{\min}, \, w_{\max}]$ and, \textbf{if} $R=1$, the Welch PSD settings (overlap percentage and $\mathcal{N}_{\mathrm{seg}}$).
\State Minimise the SG cost function $\mathcal{C}_{\mathrm{SG}}(\frac{T}{\Delta})$ in \eqref{eq:sg_cost} over the window time span $T\in[T_{{\min}}, \, T_{{\max}}]$, where $T_{{\min}}=w_{\min}\Delta$ and $T_{{\max}}=w_{\max}\Delta$,
by simulated annealing, thus obtaining the optimiser $T'$.
\If{$R=1$}
    \State Estimate the residual PSD at $T'$ for each selected Welch configuration and compute its dominant frequency.
    \State Decrease the window time span and, for each Welch configuration, retain the smallest window for which the dominant residual frequency remains equal to that of $T'$.
    \State Average the SG filter windows obtained from the different Welch configurations to obtain the final SG window time span $\hat{T}$.
\Else
    \State Set the final SG window time span $\hat{T}$ equal to $T'$.
\EndIf
\State Map $\hat{T}$ to the nearest admissible odd window size $\hat{w}$ via formula \eqref{eq:w_prime_def}.
\end{algorithmic}
\end{algorithm}

\subsection{SHADED staircase architecture for high-order derivative estimation and attractor reconstruction}
\label{sec:staircase}
SHADED is aimed at automatically obtaining successive derivative estimates from noisy measurements of \eqref{eq:generic_ode}, while preserving the signal content that is necessary for attractor reconstruction via differential embedding. A key feature of the proposed approach is the automatic estimation of all HD gains and SG window sizes from the data.

In the SHADED staircase architecture, the output of each level is used as the input to the next one, so that the estimation problem is solved in a recursive manner, as illustrated schematically by the repeated HD-SG blocks in Fig.~\ref{fig:methodology_pipeline}(c).
At each level, the proposed architecture combines an HD stage with HD gain automatically selected as in Section~\ref{sec:self-tuning_Diff} and an SG stage with window size automatically selected as in Section~\ref{sec:self-tuning_SG}: the HD stage provides the derivative information needed at each level, while the SG stage smooths the resulting sequence to suppress the high-frequency fluctuations produced by the HD step. 
It is important to note that in frequency-domain terms, the symmetry of the SG window yields an effectively zero-phase response for the smoothing operation \cite{schafer2011savitzky}, meaning that the filter does not introduce a time delay between input and output, whence the smoothed signal remains aligned with the original samples and can be passed directly to the next level without phase compensation (time lag correction).
Since the HD gain and SG window are expected to vary for different signal derivatives, the proposed approach estimates them level by level, thereby preserving the appropriate balance between convergence, oscillation suppression, and feature preservation for each estimated derivative.

Let $\ell\in [N]_0$ denote the differentiation level, where $N$ is the desired highest order of differentiation.
At $\ell=0$, the noisy measured signal $\{y_k\}_{k=0}^K$ is processed by an HD of differentiation order $n_d=0$ with tuned gain $\hat{L}^{(0)}$, yielding the output $\bigl \{z_{0,\hat{L}^{(0)},k}^{(0)}\bigr\}_{k=0}^K$, which is then processed by an SG filter with estimated window size $\hat{w}^{(0)}$, producing the denoised signal $\{u_k^{(0)}\}_{k\in\mathcal{I}^{(0)}}$ that estimates the noise-free signal $\{\bar{y}_k\}_{k\in\mathcal{I}^{(0)}}$,  where $\mathcal{I}^{(0)}\subseteq[K]_0$ denotes the set of admissible indices after removing boundary samples affected by the HD transient and the SG filter boundary effects.
The same processing is then repeated recursively. 
At level $\ell\geq 1$, the input to the block is the  signal $\{u_k^{(\ell-1)}\}_{k\in\mathcal{I}^{(\ell-1)}}$, obtained from the previous level. An HD of order $n_d=1$ is applied to $\{u_k^{(\ell-1)}\}_{k\in\mathcal{I}^{(\ell-1)}}$ while estimating the 
gain $\hat{L}^{(\ell)}$ from the input data. The first derivative component, denoted by $\bigl\{z_{1,\hat{L}^{(\ell)},k}\bigr\}_{k\in\mathcal{I}^{(\ell)}}$, is then processed by an SG filter with estimated window $\hat{w}^{(\ell)}$, yielding the smoothed derivative estimate $\{u_k^{(\ell)}\}_{k\in\mathcal{I}^{(\ell)}}$, where $\mathcal{I}^{(\ell)}\subset\mathcal{I}^{(\ell-1)}$.
Repeating this procedure up to a prescribed order $N$ gives $\bigl(\{u_k^{(0)}\}_{k\in\mathcal{I}^{(N)}},\{u_k^{(1)}\}_{k\in\mathcal{I}^{(N)}},\ldots,\{u_k^{(N)}\}_{k\in\mathcal{I}^{(N)}}\bigr)$, which can be used for attractor reconstruction, either via a time-delay embedding applied to $\{u_k^{(0)}\}_{k\in\mathcal{I}^{(N)}}$ or via a differential embedding based on all derivative estimates; see Fig.~\ref{fig:methodology_pipeline}(d) and Supplementary Fig. S17.

\section{Results}
\label{sec:results}
We demonstrate the efficacy of SHADED on multiple numerical examples, divided into three classes. First, we consider synthetic time series generated from benchmark complex dynamical systems: the chaotic Lorenz system \cite{Lorenz63} and the FitzHugh-Nagumo \cite{fitzhugh1961impulses}, the Hindmarsh-Rose \cite{hindmarsh1984model} and the Jansen-Rit \cite{jansen1995electroencephalogram} models from mathematical neuroscience. For each theoretical model, the equations, parameter values, and initial conditions are reported in Supplementary Table S1. Theoretical benchmark models allow us to evaluate SHADED under controlled conditions where the sampling period, transient removal, signal length, and noise can be specified explicitly and are known. Thus, the output of SHADED can be compared directly to a ground truth obtained from the models. Second, we consider realistic application-driven models, in the form of chaotic transistor-based circuits simulated in LTspice and inspired by the class of atypical oscillators introduced in \cite{minati2017atypical}. 
These circuits are known to generate a rich variety of chaotic behaviours \cite{Perc2005} depending on the component values. Since circuit simulations in LTspice emulate with high fidelity the functioning of analog circuits, while also allowing one to test the algorithm output against a ground truth signal, these examples bridge the gap between purely synthetic models and real-life experimental time series obtained from applications. Finally, we use SHADED to process empirical cardiovascular recordings from the University of Queensland Vital Signs Dataset \cite{liu2012university}, which provides real physiological waveforms acquired in a clinical setting; the use of this dataset allows one to relate the reconstructed attractors to physiological features that are commonly associated with in vivo blood pressure phenomena in humans. Since no ground truth attractor is available for these examples, the reconstruction must be evaluated through physiological plausibility and consistency with expected signal characteristics. 

\subsection{Attractor reconstruction for theoretical models}
\label{sec:results_models}
We consider four benchmark systems representing qualitatively different physical and biological settings: the Lorenz system, which describes thermal convection and exhibits deterministic chaos \cite{Lorenz63}; the Hindmarsh-Rose and FitzHugh-Nagumo neuronal models, which reproduce bursting and excitable or oscillatory dynamics, respectively \cite{hindmarsh1984model,fitzhugh1961impulses}; and the Jansen-Rit neural mass model, describing the collective activity of a cortical column and capable of generating alpha-band rhythms \cite{jansen1995electroencephalogram}.
For each of these models, the equations, parameter values and initial conditions are reported in Supplementary Table S1; the system trajectories are integrated in \textsc{Matlab} with the \texttt{ode45} function, using a sampling period $\Delta$ and a final time $t_{\mathrm{fin}}$.
An initial transient of duration $t_{\mathrm{tr}}$ is discarded, so that the retained sequence lies close to the attractor. 
The observations are corrupted by zero-mean white Gaussian noise that acts either as an additive perturbation, $\eta_{\mathrm{add}}$, or as a multiplicative perturbation, $\eta_{\mathrm{mult}}$ (see also Section~\ref{sec:notation} for a description of the effects of noise on the measured signal), whose variances are selected to produce challenging signal-to-noise ratios across the examples. The specific parameters used for each of the theoretical models are:
\begin{enumerate}
    \item Lorenz: $\Delta=10^{-4}$, $t_{\mathrm{fin}}=1020$, $t_{\mathrm{tr}}=1000$, additive noise $\eta_{\mathrm{add}} \sim \mathcal{N}(0,1)$,  multiplicative noise $\eta_{\mathrm{mult}} \sim \mathcal{N}(0,0.1)$.
    \item Hindmarsh-Rose: $\Delta=10^{-3}$, $t_{\mathrm{fin}}=1900$, $t_{\mathrm{tr}}=1500$, $\eta_{\mathrm{add}} \sim \mathcal{N}(0,0.1)$,  $\eta_{\mathrm{mult}} \sim \mathcal{N}(0,1)$.
    \item FitzHugh-Nagumo:  $\Delta=10^{-3}$, $t_{\mathrm{fin}}=190$, $t_{\mathrm{tr}}=40$, $\eta_{\mathrm{add}} \sim \mathcal{N}(0,0.1)$,  $\eta_{\mathrm{mult}} \sim \mathcal{N}(0,0.1)$.
    \item Jansen-Rit: $\Delta=10^{-5}$, $t_{\mathrm{fin}}=3$, $t_{\mathrm{tr}}=1.75$, $\eta_{\mathrm{add}} \sim \mathcal{N}(0,1)$,  $\eta_{\mathrm{mult}} \sim \mathcal{N}(0,0.01)$.
\end{enumerate}

The resulting noisy signal $y$ (see Supplementary Fig. S11 for the time series of all the considered noisy signals) is processed by SHADED to estimate the clean signal $\bar{y}$ and its derivatives up to order $N=2$, so that the reconstructed attractors can be visualised via differential embedding in $\mathbb{R}^3$. In these examples, the signal $\bar{y}$ is available for comparison with the output of SHADED. This allows us to study the effect of the automatic tuning on the numerical differentiation accuracy and on the resulting differential embeddings starting from the noisy signal $y$, and to compare it with the manual selection of the HD gain $L$ and the SG window size $w$ used in \cite{sutulovic2025efficient}, which was based on a highly time-consuming visual comparison of the estimated  signals with the available noise-free ground truth. To demonstrate the efficiency of SHADED, we consider three simulation settings: (i) a noise-free case that serves as the reference ground truth, (ii) a case in which $L_\mathrm{G}$ and $w_\mathrm{G}$ are guessed manually (either overestimated or underestimated), and (iii) the SHADED configuration in which $\hat{L}$ and $\hat{w}$ are directly inferred from data according to the procedures described in Section~\ref{sec:self-tuning_Diff} and Section~\ref{sec:self-tuning_SG}.

\begin{figure}[p] 
\centering\includegraphics[width=1\linewidth]{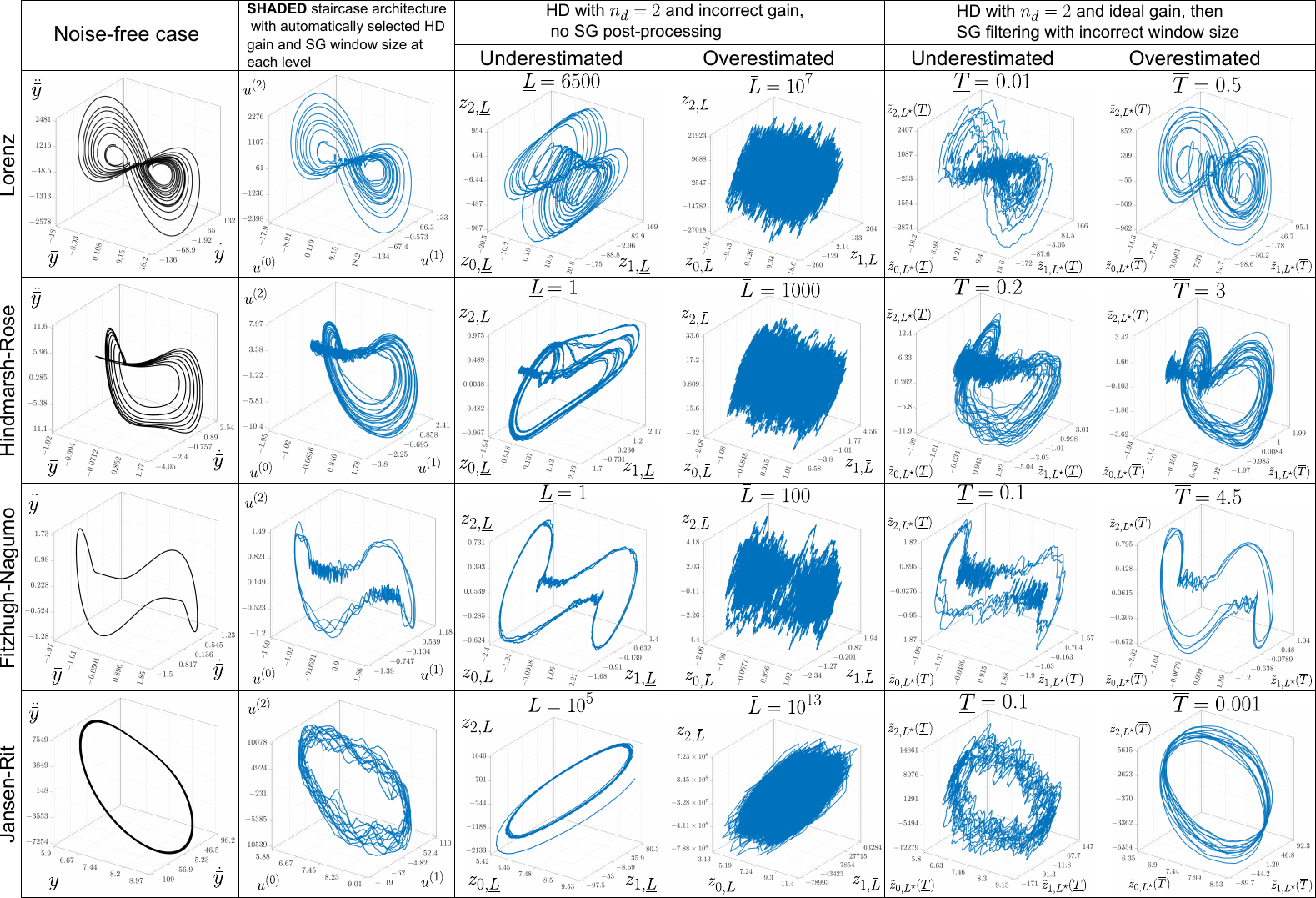}
\caption{\textbf{SHADED reproduces the noise-free differential embeddings of four benchmark systems, whereas manual HD gain or SG filter window size selection may result in severe distortions.}
Three-dimensional differential embeddings of the theoretical models discussed in Section~\ref{sec:results_models} (see Supplementary Table S1 for details), one per row. The first column shows the reference noise-free embedding computed from the ground-truth trajectory $[\bar{y}(t),\dot{\bar{y}}(t),\ddot{\bar{y}}(t)]^\top$, which is not accessible in practical applications (whence manual tuning of the HD gain $L$ and the SG window time span $T$ amounts to uninformed guessing).
The subsequent columns show the reconstructed attractors, for different parameter choices, in the additive noise setting. The results with multiplicative noise are reported in Supplementary Fig. S12.
The second column shows the attractors reconstructed by SHADED, with automatically selected HD gain and SG window: for Lorenz $\hat{L}^{(0)}=7.4\cdot 10^2,\hat{L}^{(1)}=3.2\cdot 10^5,\hat{L}^{(2)}=5\cdot10^7$ and $\hat{T}^{(0)}=0.11,\hat{T}^{(1)}=0.094,\hat{T}^{(2)}=0.11$; for Hindmarsh-Rose $\hat{L}^{(0)}=9,\hat{L}^{(1)}=8.2\cdot 10^2,\hat{L}^{(2)}=2.7\cdot 10^3$ and $\hat{T}^{(0)}=0.77,\hat{T}^{(1)}=0.9,\hat{T}^{(2)}=0.79$; for Fitzhugh-Nagumo $\hat{L}^{(0)}=6.5,\hat{L}^{(1)}=5.4\cdot 10^2,\hat{L}^{(2)}=9.5\cdot 10^3$ and $\hat{T}^{(0)}=2.3,\hat{T}^{(1)}=0.81,\hat{T}^{(2)}=0.57$; for Jansen-Rit $\hat{L}^{(0)}=3.3\cdot 10^3,\hat{L}^{(1)}=4.9\cdot 10^6,\hat{L}^{(2)}=1.7\cdot 10^9$ and $\hat{T}^{(0)}=0.024,\hat{T}^{(1)}=0.017,\hat{T}^{(2)}=0.0084$.
Despite the noise, such attractors remain close to the (inaccessible) noise-free attractors across all considered examples.
The third column shows the effect of an incorrect HD gain $L$, when only a single HD with $n_d=2$ is applied, without subsequent SG filtering; the left and right panels correspond, respectively, to the underestimated $\underline{L}$ and overestimated $\overline{L}$ values with respect to the RMSE-optimal value $L^\star$.
In the fourth column, the HD has $n_d=2$ and optimal gain $L^\star$, so as to focus on the effect of the subsequent SG filtering, applied identically to all HD outputs, with window time span that is either underestimated ($\underline{T}$) or overestimated ($\overline{T}$).}
\label{fig:results_models}
\end{figure}

The results are given in Fig.~\ref{fig:results_models}, which shows the attractors reconstructed via differential embedding for the four considered models (in the rows) in the noise-free case (first column) and then, in the case of additive noise, by SHADED, with automatically determined HD and SG parameters, including the frequency-persistent adjustment (second column), by a single HD with $n_d=2$ and either under- or over-estimated gain and no SG post-processing (third column), and by a single HD with $n_d=2$ and RMSE-optimal gain $L^\star$ followed by SG post-processing, identical for all HD outputs, with either under- or over-estimated SG window size (fourth column).

When the HD gain $L$ is underestimated, the HD estimates fail to converge properly and the reconstructed differential attractors suffer from severe geometric distortions: distinct regions of the original attractor become folded onto each other and local curvature is strongly underestimated (third column, left panels in Fig.~\ref{fig:results_models}). Conversely, when $L$ is overestimated, the HD becomes too sensitive to measurement noise $\eta(t)$ and discretisation, producing chattering components in the output signals $u^{(\ell)}$, $\ell \in [N]_0$, that completely destroy any meaningful data carried by the signal, as well as the structure of the attractor (third column, right panels in Fig.~\ref{fig:results_models} for $N=2$). To assess the effect of a poorly estimated SG window size $w$ on the signal estimates, we fix the HD gain to be equal to the optimal $L^\star$ (inaccessible in practical applications) that minimises the RMSE of the zeroth-order derivative HD estimate with respect to the noise-free signal $\bar{y}$. When $w$ is underestimated, the residual chattering persists in all derivative estimates and the resulting differential embeddings appear as a smearing of the noise-free attractor, with local crossing of nearby trajectories, especially for coordinates related to higher-order derivatives (fourth column, left panels in Fig.~\ref{fig:results_models}). 
On the other hand, an overestimated $w$ applies 
a strong low-pass filter to the estimated derivatives. This results in an overly smoothed reconstructed attractor, especially in regions that correspond to oscillations of the trajectories (fourth column, right panels in Fig.~\ref{fig:results_models}). 
Such distortion can bias both qualitative interpretations (dynamics that look simpler, because oversmoothing has removed fine-scale features) and quantitative measures (underestimation of correlation dimension, loss of fine-scale structure in recurrence plots, etc.). These simulations also show that, in the absence of the noise-free reference embedding obtained from $(\bar{y},\dot{\bar{y}},\ddot{\bar{y}})$ (first column in Fig.~\ref{fig:results_models}), which is indeed inaccessible in real applications, choosing $L$ and $w$ by hand is extremely challenging and can produce markedly distorted reconstructions.

Fig.~\ref{fig:results_models} (second column) shows the attractors reconstructed by SHADED, where the HD gain $\hat{L}$ and the SG filter window size $\hat{w}$ are automatically selected directly from the noisy measurements. Supplementary Figs. S13-S15 show the corresponding error colourmaps and distributions. 
SHADED yields differential embeddings that are visually close to the inaccessible, ideal noise-free attractor in all the theoretical examples considered. 
These results further underline that HD and SG filtering are complementary: the HD alone, even with an optimally tuned gain, can still suffer from chattering when implemented in discrete time, while the SG filter alone, which can be used to compute numerical derivatives by differentiating the local polynomial fits, cannot recover higher-order derivatives from the noisy signal with the desired accuracy \cite{schmid2022and} when used without prior HD-based differentiation. 
Their combination in our automatic procedure achieves reconstruction that is robust to noise and faithful to the underlying noise-free dynamics.

In addition to differential embeddings, we also reconstruct attractors via time-delay embedding from the denoised zeroth-order signal $u^{(0)}$ produced by SHADED, as shown in Supplementary Fig. S17. A fully automatic delay-embedding method can be obtained by feeding $u^{(0)}$ to the PECUZAL approach \cite{kramer2021unified}, which selects non-uniform delays and the embedding dimension in a fully data-driven way: the signal considered in Supplementary Fig. S17 is the output of the Lorenz system, and in both the additive and multiplicative noise scenarios the time-delay embedding obtained from $u^{(0)}$ in this way is consistent with the expected attractor geometry for this model.
By contrast, applying PECUZAL directly to the raw noisy time series $y$, without the HD-SG pre-processing, leads to an incorrect estimate of the embedding dimension, so that no meaningful attractor can be recovered. These observations indicate that SHADED serves as an essential pre-processing stage for automatic time-delay embedding techniques such as PECUZAL in the presence of noise: HD-SG delivers robust automatic denoising at the level of the scalar observable, and PECUZAL then handles the data-driven selection of delays and time-delay embedding dimension, thus yielding, overall, a completely automatic time-delay embedding methodology.

While the method in \cite{kramer2021unified} operates purely based on the raw observations in a time-delay-embedding framework,
denoised derivative estimates are an explicit output of SHADED and can be used for differential embeddings, which are preferable in many applications.

\subsection{Attractor reconstruction in noisy transistor-based chaotic circuits}
\label{sec:results_LTspice}

\begin{figure}[p] 
\centering\includegraphics[width=0.65\linewidth]{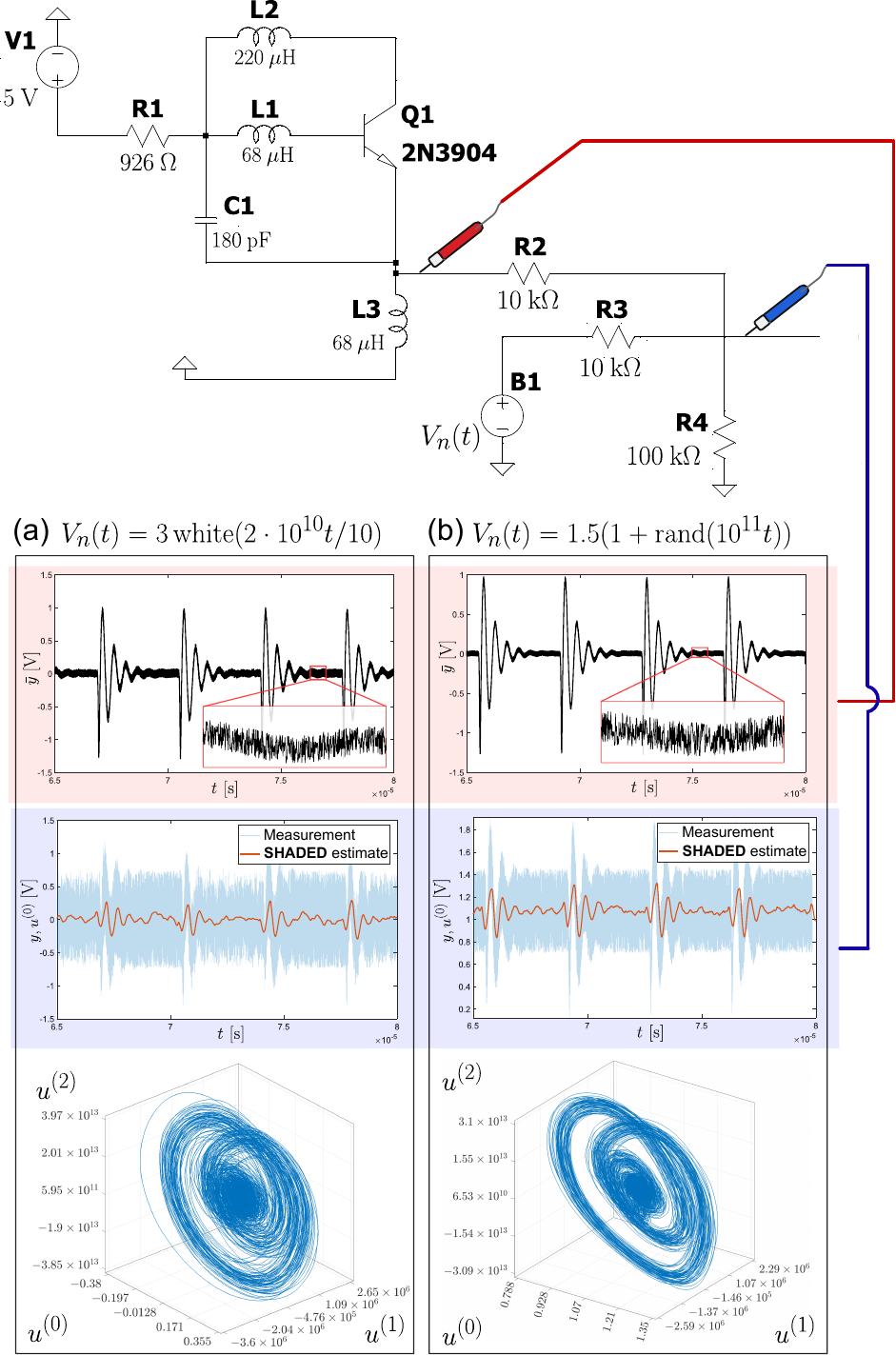}
\caption{\textbf{SHADED recovers coherent differential embeddings from a circuit with internally injected noise.}
Attractor reconstruction from the noisy transistor-based chaotic circuit considered in Section~\ref{sec:results_LTspice}. 
Top: schematic of the circuit modelled in LTspice. 
Bottom: voltage $\bar y$ before the noise source (red probe), voltage $y$ after the noise source (blue probe) along with its estimate $u^{(0)}$ generated by SHADED, and three-dimensional differential embedding obtained by applying SHADED to the noisy signal $y$, thus obtaining $u^{(0)}$ as well as its subsequent derivatives $u^{(1)}$ and $u^{(2)}$.
In (a), the rapidly fluctuating noise process $V_n(t) = 3\,\texttt{white}(2\cdot10^{10} t / 10)$ approximates a broadband, quasi-Gaussian excitation with strong high-frequency content. 
In (b), the noise process $V_n(t) = 1.5(1 + \texttt{rand}(10^{11} t))$ produces bounded, piecewise-constant fluctuations with jump-like amplitude variations. No noise-free attractor is accessible, as circuit dynamics are fully influenced by noise due to back-propagation (see zoomed-in insets in the plots of $\bar y$).}
\label{fig:results_LTspice}
\end{figure}

Reconstructing attractors of chaotic circuits and systems is of particular interest \cite{Perc2005}.
We obtain the chaotic circuit in Fig.~\ref{fig:results_LTspice}
by combining, via the LTspice simulator, the oscillator labelled as Circuit 9 in \cite[Fig. 4]{minati2017atypical} together with a noise source term. We consider two noise types that are mixed with the output of the oscillator: the first stems from a rapidly fluctuating white-noise  process of the form $V_n(t)=3\,\texttt{white}(2\cdot10^{10}t/10)$, where $\texttt{white}(\cdot)$ is the LTspice pseudo-random waveform function used to generate a smoothly varying white-noise excitation; the second stems from a bounded random process of the form $V_n(t)=1.5\bigl(1+\texttt{rand}(10^{11}t)\bigr)$, where $\texttt{rand}(\cdot)$ denotes a slowly updated random sequence that produces piecewise-constant fluctuations with jump-like amplitude variations. 
In both settings, the sampling period is $\Delta=10^{-9}$.

These simulations bridge the gap between the theoretical models in Section~\ref{sec:results_models} and the real-life data processed in the next section. 
In fact, unlike the phenomenological ODE models of Section~\ref{sec:results_models}, which are low-dimensional and prescribed directly in terms of state variables, the LTspice simulator solves the full set of circuit equations dictated by Kirchhoff's laws and the device characteristics of the transistors, capacitors, and resistors, thereby implementing a physics-based description of the oscillator rather than a reduced-order phenomenological one. 
Moreover, the noise source is embedded inside the circuit, rather than added \textit{ex post} to the simulated output; as a consequence, the noise back-propagates into the circuit equations that govern the oscillator, thereby altering the internal state evolution of the oscillator. Thus, the noisy voltage at the measurement node cannot be modelled as either purely additive or purely multiplicative noise, whence no exact ground truth is available in these simulations, although the LTspice simulator still allows one to obtain a reference voltage of the oscillator alone. Therefore, differently from the examples of the previous section, where the ground truth was available, in these examples manual tuning of the HD gain and the SG window is expected to produce unreliable attractor reconstructions.

Moreover, since the noise back-propagates into the oscillator's internal dynamics, the attractor reported in \cite{minati2017atypical} is no longer accessible, and a direct quantitative comparison with those time-delay-based reconstructions is not meaningful.
Our objective is therefore to assess whether SHADED can recover a geometrically coherent differential embedding from measurements generated by the noise-perturbed transistor network.
Since no ground truth is available, the assessment is necessarily qualitative and is based on three geometric features that a faithful reconstruction is expected to display, and which we define here once and use throughout this section.
First, the trajectory should concentrate on thin sheets, that is, successive passages of the trajectory through the same region should remain visually distinguishable as separate thin layers rather than merging into a thick band.
This reflects the fact that a chaotic attractor of a dissipative system often occupies a set of zero volume in the embedding space \cite{ott1993chaos}; in our case, any apparent thickness is attributable to noise or to residual estimation error rather than to the dynamics.
Second, the embedding should display a coherent recurrent organisation, meaning that repeated trajectory passages follow the same ordered circulation through the same region of the embedding space: successive revolutions should have comparable orientation, magnitude, and nesting, and should remain concentrated around a common centre rather than appearing as isolated, irregularly oriented paths distributed throughout the embedding space.
Unlike the Lorenz system in Fig.~\ref{fig:results_models}, whose attractor is organised in two distinct “wings”, the circuit considered here produces a “single-scroll” structure: the trajectory circulates around one centre along nested loops of different amplitudes, which form a banded annular region with a densely visited core, so the relevant feature is the persistence of this nesting.
Third, these features should be reproducible, in the sense that the same arrangement of sheets and the same nested circulation are recovered under both noise types described above and across noise intensities.
A reconstruction that changes qualitatively when the noise realisation or its amplitude is varied is instead dominated by the noise rather than by the underlying dynamics.

Empirically, the frequency persistence criterion yields noticeably better reconstructions than minimising the SG cost $\mathcal{C}_{\mathrm{SG}}(w)$ alone.
We hypothesise that, in the considered transistor-based circuit, the HD output contains intense high-frequency chattering induced by the internal noise back-propagation, so the chattering-suppression term in $\mathcal{C}_{\mathrm{SG}}(w)$ remains large over a wide range of window sizes.
As a result, the cost-minimising window is shifted toward relatively large values of $w$, because only such wide windows reduce the chattering term sufficiently to compensate for the penalty from signal smoothing.
This oversmooths the lower-frequency modulations that carry information about the oscillator's state evolution, causing the nested loops of the reconstructed attractor to collapse onto each other, so that the embedding degenerates into a rounded object in which individual passages can no longer be distinguished (see Supplementary Figs. S18 and S19).
By contrast, the persistence-based adjustment (as shown in Fig.~\ref{fig:results_LTspice}) selects a smaller window that still removes the chattering band but preserves the lower-frequency content, yielding an attractor in which loops of different amplitude remain separated and the derivative coordinates $u^{(0)}$, $u^{(1)}$ and $u^{(2)}$ retain a substantially wider range; for instance, the range of $u^{(2)}$ is larger by a factor between roughly $3.5$ and $7$, depending on the noise type.
As mentioned above, we prefer to tolerate some residual chattering rather than to lose the fine geometric features of the attractor.
An additional advantage of the persistence criterion concerns the second noise source, $V_n(t)=1.5\bigl(1+\texttt{rand}(10^{11}t)\bigr)$, which is piecewise constant with jump-like amplitude variations: its energy is not delivered uniformly in time but in short, irregularly spaced episodes, as each jump produces a transient surge of the chattering amplitude in the HD output.
A window selected by minimising $\mathcal{C}_{\mathrm{SG}}(w)$, which aggregates the signal variance over the whole record, is sensitive to these episodes, whereas the persistence criterion only requires that the dominant frequency of the SG residual be unchanged and is therefore largely unaffected by variations of its amplitude.
Consequently, the resulting attractor retains its nested banded structure across both noise types and over a range of noise intensities (see Fig.~\ref{fig:results_LTspice} and Supplementary Figs. S20 and S21).

Our assessment therefore relies on visual inspection of the reconstructed three-dimensional attractor in differential coordinates, using the three features introduced above.
The key observation is that, after automatic HD gain selection and SG window tuning with the frequency persistence adjustment, the embedding occupies a bounded annular region of the $(u^{(0)},u^{(1)},u^{(2)})$ space and displays a reproducible organisation: the trajectory circulates around a single centre along nested loops whose amplitude varies from one revolution to the next, the loops remain individually distinguishable instead of merging, and the innermost part of the region is visited frequently, so that the object does not degenerate into an amorphous cloud filling the available volume.
The two-dimensional projections, reported for the random-step and white-noise configurations in Supplementary Figs. S20 and S21, make this structure easier to inspect: in the $(u^{(0)},u^{(1)})$ and $(u^{(1)},u^{(2)})$ planes, nested circulation is visible; whereas, in the $(u^{(0)},u^{(2)})$ plane, a collapse onto a thin, nearly straight band indicates that the second derivative estimate remains tightly related to the signal itself, as expected for predominantly oscillatory dynamics, while the thinness of this band provides a direct visual indication of the residual noise retained in the derivative estimates.
Note that each panel of the supplementary figures is drawn with its own axis limits, so the comparison between the two window-selection settings should be based on the reported coordinate ranges rather than on the apparent thickness of the trajectories.
Such features are difficult to obtain if either the HD gain or the SG window is strongly mis-tuned, especially when the noise acts within the circuit equations: an underestimated gain folds distinct portions of the attractor onto each other, an excessive gain fills the volume with chattering, and an oversized SG window collapses the nested loops into a single rounded shape, exactly as observed for the theoretical models in Fig.~\ref{fig:results_models}.
The presence of such features therefore provides indirect evidence that SHADED extracts a differential-embedding representation reflecting the dominant chaotic regime of the noisy circuit.

\subsection{Attractor reconstruction from cardiac empirical data}
\label{sec:results_data}

We apply SHADED to empirical cardiovascular signals from the University of Queensland Vital Signs Dataset, which contains waveform and numeric recordings acquired during anaesthesia for 32 surgical cases at the Royal Adelaide Hospital \cite{liu2012university}. The relevant cardiovascular channels are the photoplethysmographic (PPG) waveform and the arterial blood pressure (ABP) waveform, both sampled with a temporal resolution of $10$ ms.
The PPG signal provides a noninvasive optical measure of peripheral blood-volume changes, whereas the ABP waveform reflects the pulsatile pressure generated by the heart through an invasive arterial catheter.

We consider three patients (datasets 05, 26, and 27) for which both PPG (channel “pleth”) and ABP (channel “ART”) recordings are available over overlapping time intervals, and we extract segments that contain typical cardiac cycles together with baseline drifts associated with changes in the patient’s activity, such as repositioning, movement, coughing, or changes in breathing pattern \cite{aston2018beyond}.
These slow drifts translate the entire reconstructed attractor in the derivative-coordinate space, leaving its shape essentially unchanged (for instance, see the leftmost panel in the lowest row of Supplementary Fig. S27).
Time is expressed in seconds and SHADED is applied to each waveform to estimate successive derivatives up to order $N=2$, thereby yielding the coordinates $u^{(0)}$, $u^{(1)}$, and $u^{(2)}$ used to construct differential embeddings.
For the SG filter, we set the maximum window size to $w_{\max}=101$ samples, which corresponds to a window of approximately $1$ s and is therefore comparable to the characteristic time scale of the cardiac cycle.
This choice illustrates how prior knowledge about the underlying physiology (including, e.g., about the dominant time scales) can be incorporated as a manual constraint, while the remaining parameters are determined automatically by the HD gain-selection and SG-tuning procedures described, respectively, in Section~\ref{sec:self-tuning_Diff} and Section~\ref{sec:self-tuning_SG}, and exploited to set bounds ensuring that automatic tuning operates within a physiologically meaningful range, and yields embeddings whose
attractor geometry aligns with canonical features of the cardiac cycle.

\begin{figure}[p]
\centering
\includegraphics[width=1\linewidth]{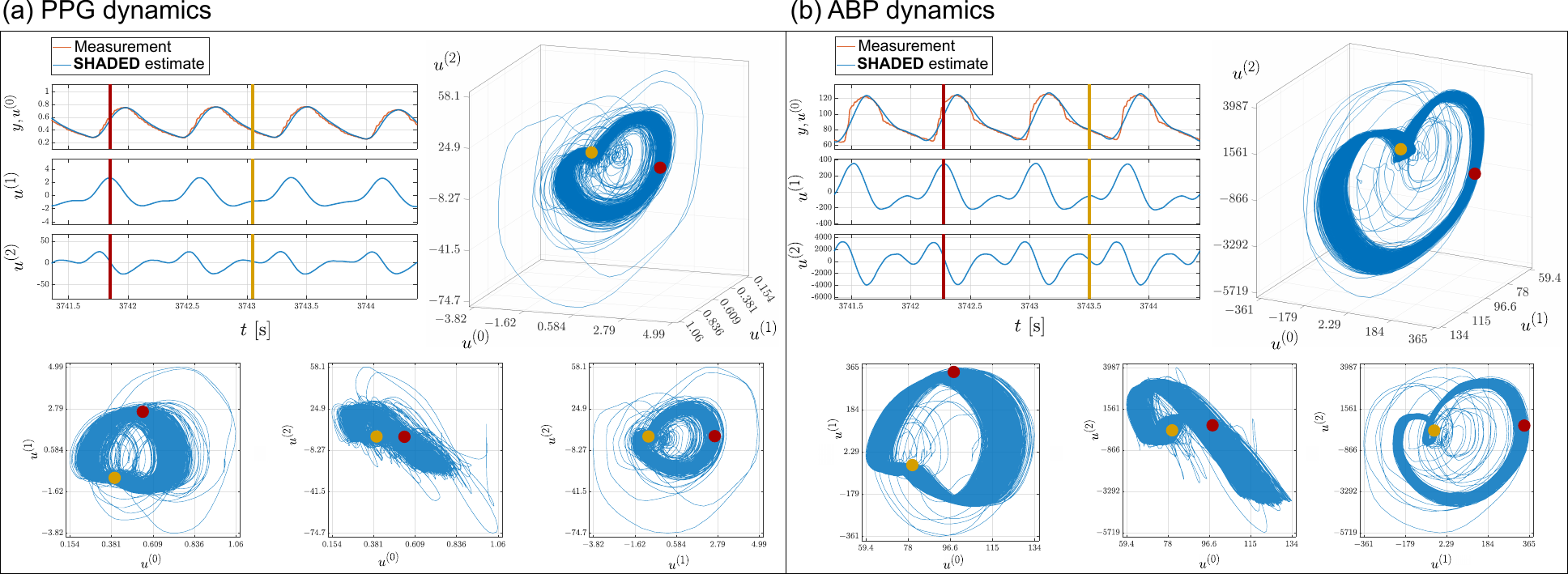}
\caption{\textbf{SHADED reconstructs differential embeddings from real PPG and ABP recordings, revealing a double-loop structure that may be associated with distinct landmarks of the cardiac waveform.}
Differential embeddings of cardiovascular waveforms from recording \textsc{uq\_vsd\_case26\_fulldata\_07} 
\cite{liu2012university}.
(a) PPG dynamics (time series “pleth”) and (b) ABP dynamics (time series “ART”). Both series are recorded over the same time interval.
Each panel contains the time series for the zeroth-, first-, and second-order derivative estimates, with the zeroth-order plot also showing the input signal $y$, as well as the three-dimensional differential embedding and its two-dimensional projections.
The full SHADED processing of both signals, including automatic gain and window selection up to second-order derivatives with the frequency-persistence criterion enabled, requires approximately \textbf{50 seconds} on a Dell Inspiron 16 laptop with 16 GB RAM and 1.90 GHz Intel core i5-1340P processor running Windows, and is thus substantially faster than applying the Schreiber-Grassberger denoising method \cite{grassberger1993noise} to time-delay embeddings of comparable length and dimension \cite{sutulovic2025efficient}, which, based on the numerical tests we conducted, would require between one and two hours of computation.
Two vertical lines in the time series (shown in different colours) mark the instants corresponding to two consecutive peaks in the first derivative: the $u$-peak in red and the $w$-peak in gold, as defined in \cite{suboh2022analysis}.
The corresponding points on the reconstructed trajectory are highlighted in the differential embeddings and projections, thereby linking specific waveform landmarks to their geometric location on the attractor.
In particular, the first peak ($u$-peak) lies on the outer loop of the double-loop structure, while the second peak ($w$-peak) lies on the inner loop, reflecting their association with distinct phases of the cardiac cycle. These features appear to be recurrent across patients.}
\label{fig:cardio_differential_embeddings}
\end{figure}

Fig.~\ref{fig:cardio_differential_embeddings} reports, for one representative patient, the PPG (left) and ABP (right) waveforms along with their corresponding differential embeddings. 
In each case, the top plots show the time series associated with the zeroth-, first-, and second-order estimates together with the three-dimensional embedding, while the bottom plots show all two-dimensional projections, so that the temporal morphology and the geometric structure in derivative coordinates can be inspected jointly. 
Two vertical lines in the time series (shown in different colours) mark the instants corresponding to two consecutive peaks in the first derivative, specifically the $u$-peak and $w$-peak as defined in \cite{suboh2022analysis}; the corresponding points on the reconstructed trajectory are highlighted in the differential embeddings, thereby linking specific waveform landmarks to their geometric location on the attractor.
Analogous embeddings for the other two patients, shown in Supplementary Section S2.3, confirm that the identified qualitative geometric features are preserved across different recordings.

For all three patients, the $(u^{(1)},u^{(2)})$ projection of the ABP embeddings consistently exhibits a double-loop structure consisting of one larger outer loop that folds into a smaller inner loop.
The PPG embeddings share the same overall organisation (as expected, since they reflect the same underlying cardiac cycle), but the double loop is less sharply expressed: the inner loop is clearly visible only in part of the considered segments (see Supplementary Figs. S31-S33), whereas in the remaining ones the trajectories mainly form a broad ring-shaped region around a central area and the passages associated with the inner loop overlap with the surrounding trajectories rather than forming a separate, clearly distinguishable branch.
This difference is consistent with the fact that PPG is an indirect optical measure of blood-volume changes rather than a direct pressure measurement, so that its waveform features are less distinct than in ABP \cite{aston2018beyond,horandtner2022attractor}.

We hypothesise that these two loops correspond to two main phases of the cardiac cycle: the rapid systolic upstroke and systolic peak, and the return branch of the cardiac cycle including the dicrotic notch and subsequent diastolic decay \cite{pal2024algorithm}. 
The plausibility of this interpretation may depend on recording conditions, patient specific cardiovascular properties, and the presence of artifacts or irregular beats; we therefore regard the geometric association between loop size and specific waveform phases as a plausible but not yet formally validated description, which is left for future investigation. 

In this context, differential embeddings may be particularly useful because derivatives accentuate local changes in slope and curvature, which can make landmarks such as the dicrotic notch easier to localise than in time-delay embeddings, which preserve the attractor only up to a homeomorphism, and thereby do not provide explicit information about derivatives. 
More specifically, in arterial pressure waveforms the dicrotic notch is typically associated with a localised change in curvature of the pressure waveform \cite{suboh2022analysis,pal2024algorithm}. 
In derivative coordinates, such curvature changes would be expected to produce pronounced excursions in $u^{(2)}$ relative to the surrounding trajectory, so that the notch and related features could, in principle, become encoded in the geometry of the inner loop and be distinguished from smoother portions of the cardiac cycle even when the raw waveform is affected by baseline wander or amplitude variability. This suggests that derivative-based embeddings have the potential to separate morphologically distinct phases of the cardiac cycle in a way that may be less sensitive to slow drifts and amplitude scaling than classical time-delay embeddings, though a quantitative validation of this advantage, and a direct identification of the dicrotic notch in the present recordings, are left to future work.

The SG window-size selection plays an important role in the clarity of the reconstructed attractor. 
In the cardiac recordings considered here, minimising the SG cost $\mathcal{C}_{\mathrm{SG}}(w)$ alone tends to favour relatively large windows that strongly smooth the HD output.
As a consequence, the reconstructed trajectory occupies wider bands in derivative-coordinate space, and the inner and outer loops can overlap rather than remain clearly separated.
The resulting loss of loop separation makes the two-loop organisation more difficult to identify (see Supplementary Figs. S22-S24). 
In contrast, the additional frequency-persistence criterion introduced in Section~\ref{sec:freq_persistence}, applied within the manually imposed bound $w\leq w_{\max}=101$ samples, systematically selects smaller windows that remove high-frequency chattering while preserving the low-frequency content associated with the heartbeat dynamics.
The resulting embeddings retain a substantially wider range of $u^{(2)}$, typically larger by a factor between $2$ and $7$, and in most ABP recordings display more sharply defined loops with a clearer separation between the inner and outer branches, so that the double-loop pattern is easier to identify (see Supplementary Figs. S25-S27).
For PPG, the additional content recovered in $u^{(2)}$ is only partly informative: the inner branch is resolved in fewer segments and part of the range expansion is due to isolated irregular beats and drift episodes (see Supplementary Figs. S31-S33), consistently with PPG being an indirect optical measure whose waveform lacks the localised curvature features of the arterial pressure waveform.
In a few segments the smaller window also lets through part of the residual chattering, which is the trade-off already discussed in Section~\ref{sec:freq_persistence}.

Processing multiple patients also highlights that the automatically tuned parameters adapt to the characteristics of each recording, rather than taking a universal fixed value. 
The selected gains and windows can differ between patients and between PPG and ABP signals, which supports the need for data-driven tuning and indicates that SHADED can accommodate inter-patient variability without manual reconfiguration of its internal parameters. 
The examples we show support the hypothesis that automatically tuned differential embeddings can serve as useful representations for cardiovascular waveforms, in addition to existing time-delay-coordinate attractor analyses \cite{nandi2018novel,aston2018beyond,horandtner2022attractor}.
Moreover, the double-loop structure found in the $(u^{(1)},u^{(2)})$ plane, which is consistent across patients in the ABP recordings, makes these embeddings particularly promising as inputs for downstream geometric and topological analyses, such as persistent homology or other attractor-based feature-extraction techniques, which require the main geometric features of the reconstruction to be consistently recovered across recordings.

Ultimately, SHADED allows us to construct derivative-based embeddings without manual tuning of differentiator gains or SG filter smoothing windows for each recording, which is advantageous in settings where many patient records must be processed systematically and the noise properties and waveform morphology vary widely. 
The results in Fig.~\ref{fig:cardio_differential_embeddings} further suggest that differential embeddings obtained via SHADED can provide robust geometric features useful for subsequent processing tasks such as beat detection, signal-quality assessment, and classification.
The need for robust derivative estimation is further illustrated by the finite-difference baseline in Supplementary Fig. S34: direct finite differences applied to the LTspice, ABP, and PPG signals strongly amplify noise and do not recover the structured, reproducible attractor geometry obtained with SHADED.

\section{Conclusions, practical considerations and limitations}
\label{sec:conclusions}

This work addressed the data-driven and automatic reconstruction of dynamical attractors from noisy scalar time series through differential embeddings.  
We proposed an automatic staircase methodology, referred to as SHADED (\textbf{S}avitzky-Golay and \textbf{H}omogeneous-differentiator based \textbf{A}utomatic \textbf{DE}noising and \textbf{D}ifferentiation), that combines HD with SG filtering and infers the differentiation and filter parameters directly from the data.
At each staircase level, the HD gain is inferred from the statistics of the differentiator residual, while the SG window size is chosen to balance chattering suppression against preservation of the low-frequency signal content, thereby producing derivative estimates that can be used to build differential embeddings without manual parameter guessing.
Across theoretical dynamical models of chaotic and neuronal systems, the resulting embeddings were consistently faithful to the reference attractors, indicating that the proposed automatic parameter inference from data achieves reconstructions that remain close to the ideal noise-free baselines in the tested scenarios. Also for noisy transistor-based chaotic circuits and real-life cardiovascular recordings, where no noise-free ground truth is available,  differential embeddings with a coherent structure were obtained, which reveal interesting features of the dynamics.

At the same time, SHADED is not to be regarded as a universal solution. 
It works best when the underlying noise-free signal is reasonably smooth and sampled with sufficient temporal resolution, as in the case of cardiovascular data.
By contrast, for low-resolution or sparsely sampled records, such as some climate time series, reliable derivative estimation and attractor reconstruction become much harder, because the available measurements may not contain enough temporal detail to capture the relevant dynamical scales \cite{lekscha2018phase}.
In such cases, reconstruction is challenging for any method, since coarse sampling limits the information available about the underlying dynamics.

A further practical limitation is that each HD-SG level removes samples near the boundaries, so the usable segment becomes shorter as the derivative order increases, which prevents the methodology from successfully handling extremely short time series. 
This limitation is also common in time-delay embedding, which likewise requires a minimum recording length to yield reliable reconstructions \cite{bradley2015nonlinear}. 
In the synthetic and cardiovascular examples considered in Section~\ref{sec:results_models} and Section~\ref{sec:results_data}, however, the automatically selected SG windows remain moderate, and the staircase consistently yields derivatives up to order $N=2$ without exhausting the available data, suggesting that recordings of typical length in these applications are sufficient for faithful attractor reconstruction.

Also, since the architecture is recursive, errors introduced at one level can propagate to subsequent levels. 
This is a difficulty shared by other iterative denoising schemes, such as the Schreiber-Grassberger method, whose repeated local averaging can be affected by noise and may distort the final attractor estimate \cite{grassberger1993noise}.
SHADED mitigates error accumulation by estimating HD gains and SG window sizes separately at each level, and by allowing optional manual adjustment when prior information on the dynamics or noise is available.
In the numerical experiments this level-wise tuning preserves the geometry of the reconstructed attractor across derivative orders even when the noise is strong, as illustrated in Fig.~\ref{fig:results_models} and Fig.~\ref{fig:results_LTspice}. 
Strong baseline drifts, abrupt artifacts, and severe sampling irregularity fall outside the intended differentiator framework and can still degrade performance; in such cases, it is important to combine SHADED with other denoising techniques whenever more information about the underlying system is available, whereas in the examples studied here the recordings were suitable for the staircase to operate without additional preprocessing. 
For non-uniformly sampled data, discrete-time HD can be implemented directly on variable sampling grids \cite{levant2020robust}, and SG windows can be defined via time neighbourhoods rather than fixed numbers of samples, so that each local polynomial fit uses all measurements within a prescribed time interval around the point of interest.
In our implementation, the SG window size is inferred from data by also analysing the PSD of the SG residual, which can be estimated from the irregularly sampled data using the Lomb-Scargle algorithm \cite{press1989fast}. 
As with other phase-space reconstruction schemes for non-uniformly sampled noisy time series \cite{lekscha2018phase}, the sampling pattern, and especially the presence of large sampling gaps, strongly influences the reconstruction quality, so the method should be applied with particular care when only sparse or highly uneven records are available. 
Finally, the residual-based costs used to tune HD and SG are designed to operate across a broad range of noise levels and signal morphologies, but, like other heuristic parameter-selection criteria, they are not universally optimal and may require adaptation when noise statistics or signal structure differ substantially from the cases considered in this paper; nevertheless, the simulations reported here indicate that, for the neuronal, electronic, and cardiovascular systems studied, the chosen costs yield derivative estimates and reconstructed attractors that either remain close to their noise-free counterparts, when these are available, or exhibit a coherent structure that is consistent with prior information about the dynamics and, at the same time, reveals novel and interesting features.

Several directions for further work follow naturally. One is to design residual-based costs and optimisation strategies that are better suited to particular classes of dynamical systems and to analyse more systematically how different choices of cost function affect reconstruction quality and derivative accuracy.
Another direction is to develop adaptive criteria for choosing the staircase depth based on data-driven estimates of the embedding dimension, so that the recursion stops once the reconstructed coordinates appear sufficient for a faithful representation of the attractor and additional derivative levels no longer provide a meaningful improvement.
It would also be valuable to study more systematically how the reconstructed attractors interact with downstream methods using them to forecast, classify, or detect changes in the underlying dynamics, so as to quantify how automatic differential embeddings influence task performance.
More broadly, the results support the view that carefully designed numerical differentiation, when coupled with attractor reconstruction, provides a reliable route to automatically characterising the geometry of nonlinear dynamics from noisy time series.

\section*{CRediT authorship contribution statement}
\textbf{U.S.:}
conceptualisation,
data curation,
formal analysis,
investigation,
methodology,
software,
visualisation,
writing - original draft,
writing - review and editing.

\textbf{D.P.:}
conceptualisation,
methodology,
visualisation,
writing - review and editing.

\textbf{R.K.:}
conceptualisation,
funding acquisition,
formal analysis,
methodology,
supervision,
validation,
visualisation,
writing - original draft,
writing - review and editing.

\textbf{G.G.:}
conceptualisation,
funding acquisition,
methodology,
project administration,
supervision,
visualisation,
writing - review and editing.

\section*{Code availability}
The \textsc{Matlab} code to reproduce all our results as well as the Supplementary Material is publicly available on GitHub at
{\texttt{https://github.com/Uros-S/SHADED}} .

\section*{Data availability}
All the data used to support the findings of this study either are included in the article, or can be generated using the code provided in the cited GitHub repository, or are publicly available from the University of Queensland Vital Signs Dataset \cite{liu2012university}.

\section*{Funding}
Work supported by the European Union through the ERC INSPIRE grant (project number 101076926). Views and opinions expressed are however those of the authors only and do not necessarily reflect those of the European Union or the European Research Council Executive Agency. Neither the European Union nor the European Research Council Executive Agency can be held responsible for them. R. Katz also acknowledges support from the Alon Fellowship, awarded by the Council of Higher Education of Israel.

\section*{Declaration of competing interest}
The authors declare that they have no known competing financial interests or personal relationships that could have appeared to influence the work reported in this paper.

\section*{Acknowledgements}
The authors are grateful to Arie Levant for valuable discussions on the theory and applications of Homogeneous Differentiators, which were instrumental in inspiring the work in \cite{sutulovic2025efficient} and the present study.

 \bibliographystyle{elsarticle-num} 
 \bibliography{references}






\end{document}